\documentclass[aps,prd,preprintnumbers,superscriptaddress]{revtex4}
\usepackage{amsmath}
\usepackage{amssymb}
\usepackage{amsthm}
\usepackage{todonotes}
\usepackage{mathtools}
\usepackage{mathrsfs}
\usepackage{bm}
\usepackage{slashed} 
\usepackage{graphicx}
\usepackage{multirow}
\usepackage{tikz}
\usepackage[caption=false]{subfig}
\usepackage{relsize}	
\usepackage{array}
\usepackage{float}
\usepackage{color}
\usepackage{xcolor}
\usepackage{soul}
\usepackage{verbatim} 
\usepackage{hyperref}
\usepackage{pifont}

\allowdisplaybreaks
\newcommand{\beq}{\begin{equation}}
	\newcommand{\eeq}{\end{equation}}
\newcommand{\be}{\begin{eqnarray}}
	\newcommand{\ee}{\end{eqnarray}}

\usepackage[normalem]{ulem}

\newcommand{\bsq}{{\boldsymbol q}^{\perp}}

\newcommand{\bsk}{{\boldsymbol k}^{\perp}}

\newcommand{\bska}{{\boldsymbol\kappa}^\perp}
\newcommand{\bskapr}{{\boldsymbol\kappa}^{\prime \perp}}
\newcommand{\bskasq}{{\kappa}^{\perp2}}
\newcommand{\bskpr}{{\boldsymbol k}^{\prime \perp }}

\newcommand{\bsb}{{\boldsymbol b}^{\perp}}

\newcommand{\bsP}{{\boldsymbol P}^{\perp}} 
 
\newcommand{\bsp}{{\boldsymbol p}^{\perp}} 
\newcommand{\bspp}{{\boldsymbol p}^{\perp 2}}

\newcommand{\es}{&=&}

\newcommand{\nnn}{\nonumber\\}

\newcommand{\bsD}{\boldsymbol{\Delta}^{\perp}}

\newcommand{\bs}{\boldsymbol}
\begin{document}

\begin{titlepage}

\title{Gluon contribution to the angular momentum distribution of a dressed quark state
}

\author{Asmita Mukherjee}
\email{asmita@phy.iitb.ac.in}
\author{Sudeep Saha}
\email{sudeepsaha@iitb.ac.in} 
\author{Ravi Singh}
\email{ravi.singh298@iitb.ac.in}
\affiliation{Department of Physics,
Indian Institute of Technology Bombay, Powai, Mumbai 400076,
India}
\begin{abstract}
{We compute the contribution of the gluonic component of the energy-momentum tensor (EMT) to the angular momentum density in various decompositions.}{ We use the light-front Hamiltonian technique, and a two-component formalism in light-front gauge, where the constrained degrees of freedom are eliminated. Instead of a nucleon, we consider a simple composite spin-$1/2$ state, namely a quark dressed with a gluon. We present two dimensional light-front distributions in transverse impact parameter space, and compare the different angular momentum decompositions at the density level. Incorporating also the contribution coming from the quark part of the EMT, we verify the spin sum rule for such a state. }
\end{abstract}
\date{\today}
\maketitle

\end{titlepage}

\section{Introduction}

Understanding the spin structure of a nucleon has been a major challenge in hadron physics. Over three decades ago, the quark model provided a simple, non-relativistic, constituent-based explanation for proton spin \cite{Fritzsch:1973pi, Close}. However, the European Muon Collaboration's deep-inelastic scattering experiments revealed that quark spin contributes only a small fraction to the proton's spin \cite{EuropeanMuon:1987isl, EuropeanMuon:1989yki, Kuhn:2008sy, Anselmino:1994gn}. With the advent of QCD, it was thought that the rest of the nucleon spin could come from the intrinsic spin of the gluons. However, even the gluon spin measured through different experiments could not account for the total proton spin \cite{deFlorian:2014yva, AGEEV200625, Boyle:2006ab, 10.1063/1.2220260, Ji:2013fga, Borsa:2024mss, Karpie:2023nyg}. These findings show that the nucleon is a highly relativistic bound state, with its spin arising largely from the interactions among its constituents, rather than solely the result of the intrinsic properties of its constituents. Therefore, a significant fraction of the spin budget has to be accounted for by the orbital angular momentum (OAM) of both quarks and gluons. Present experiments like JLab 12 GeV \cite{Dudek:2012vr} and {the upcoming} Electron-Ion Collider (EIC) at Brookhaven National Lab  \cite{AbdulKhalek:2021gbh}  aim to measure the OAM and spin of all partons with increased accuracy; this will help us understand the origin of proton spin. On the theory side, for a long time, issues like gauge invariance of the spin decomposition of the nucleon puzzled researchers. With significant advancement in the past decade, this issue now has been resolved \cite{Leader:2013jra}. The quark-gluon interactions as well as the intrinsic transverse motion of the quarks and gluons play a major role in the spin of the nucleon. In fact, the decomposition of the nucleon spin $1/2$ into orbital and intrinsic parts coming from quarks and gluons, is not unique. For a long time, Ji \cite{Ji_1997} and Jaffe-Manohar (JM)\cite{Jaffe:1989jz} decompositions were used in this context. The Jaffe-Manohar approach gives the decomposition of spin into intrinsic and orbital parts coming from quarks and gluons. This has a partonic interpretation but it is not gauge-invariant, while Ji's method is gauge-invariant but doesn't separate the contribution of the gluon into spin and orbital parts. Ji's method is an improvement upon the longest-known Belinfante decomposition \cite{Belinfante1939OnTS, BELINFANTE1940449, Rosenfeld} which only decomposes proton spin into total angular momentum of quarks and gluons, each gauge invariant.  Recent advances suggest that by introducing the physical component of the gauge field, both Ji's and JM's decompositions can be made gauge-invariant and complete, resulting in two distinct classes of decomposition, that are called canonical and kinetic respectively \cite{Chen:2008ag, Chen:2009mr, Wakamatsu:2010qj, Wakamatsu:2010cb, Hatta:2011zs,  Leader:2013jra}. This gauge-invariant extension of angular momentum decomposition is essential in relating the theoretical quantities to experimental observables. A thorough overview of these decompositions, including the related theoretical challenges and their solutions, and the latest developments are provided in \cite{Leader:2013jra, Ji:2020ena}. 

While collider experiments and theoretical developments have provided valuable insights into the origin of nucleon spin, most efforts have focused on integrated quantities, such as the total spin and the orbital angular momentum contributions of quarks and gluons. Little attention has been given to how these quantities are distributed spatially inside the nucleon. For instance, all the aforementioned decompositions differ from each other by the addition of boundary terms or what are also called superpotentials. In the integrated quantities, these do not contribute. However, the presence of superpotentials can affect the different decompositions of the nucleon spin at the density level. In particular, total angular momentum densities computed using different decompositions do not agree with each other in the absence of superpotentials \cite{Lorce:2017wkb, PhysRevD.94.114021}. So it is interesting to investigate the role of superpotentials at the density level. In the literature, spatial distributions of angular momentum are typically constructed in two different frames in order to give them a probabilistic or density interpretation. The first involves calculating 3D distributions in the Breit frame, but these require relativistic corrections unless the nucleon is assumed to be infinitely massive \cite{Polyakov:2002yz, Goeke:2007fp}. The alternative approach is to construct 2D distributions within the light-front framework, which, due to Galilean symmetry in the plane transverse to the direction of motion in the light-front approach, are free from relativistic corrections. This gives the picture of nucleon spin distribution in the impact parameter space. A phenomenological way to study the spatial distribution of angular momentum is through the generalized parton distributions (GPD) which play a vital role in the hard exclusive reactions like deeply virtual Compton scattering (DVCS) and deeply virtual meson production (DVMP) \cite{diehl2003generalized, Boffi:2007yc, Bacchetta:2016ccz, dHose:2016mda, Kumericki:2016ehc, Collins:1996fb}. GPDs do not have a probabilistic interpretation. However, when the momentum transfer in the process is purely in the transverse direction, the Fourier transform of GPDs with respect to transverse momentum transfer gives access to the impact-parameter distribution of partons. Such impact-parameter distributions have a probabilistic interpretation \cite{Burkardt:2000za} and their moments give various gravitational form factors which are essential in studying the distribution of energy, pressure, shear, and also angular momentum in nucleons \cite{Burkert:2018bqq, Burkert:2021ith, Lorce:2018egm, Polyakov:2018zvc}. Surface terms that link different decompositions of angular momentum influence the relationship between the Fourier transform of the  GPDs in impact parameter space and the angular momentum distributions in the transverse plane.
 
 Different definitions of the angular momentum distributions were investigated in \cite{PhysRevD.94.114021, Lorce:2017wkb} using a scalar-diquark model (SDQM), and it was shown that when all the surface terms are included, the total angular momentum distribution is the same for Belinfante and Ji decomposition. Distributions of quark angular momentum have also been studied in a light-front quark-diquark model using soft wall AdS/QCD \cite{Kumar:2017dbf}  and in the BLFQ framework \cite{Liu:2022fvl}. Although a few model-dependent studies have been conducted on understanding the effect of surface terms on the nucleon spin decomposition problem, they lack two important ingredients. First is that none of these models have a gluonic degree of freedom because of which contribution of the gluon to the angular momentum or its distributions cannot be calculated in these models.  Secondly, due to the absence of gluons in these models, it is not possible to verify the integrated spin sum rule for any of the decompositions, where the gluon contribution plays an important role. Only very recently, several spectator models have been proposed to explore gluon parton distributions \cite{Bacchetta:2020vty,Lyubovitskij:2020xqj,Lu:2016vqu,Chakrabarti:2023djs,PhysRevD.108.054038} but none of them have discussed the effect of surface term or superpotential term to the distribution of gluon angular momentum. This is partially due to the absence of a direct observable, unlike the quark spin density, which can be linked to the axial-vector form factor. This is not solely due to the gluon's color-gauge-variant nature, as even color-gauge-invariant quantities, like quark spin density in a world without weak interaction, would be unobservable without an external probe \cite{Wakamatsu:2019ain}. Observability depends on the existence of an external probe that couples to the quantity, not just gauge invariance. Also, in spectator-type models for the nucleon, it is very difficult to satisfy the momentum sum rule \cite{Bacchetta:2008af} as well as the spin sum rule \cite{Gurjar:2021dyv}. This is because in most spectator-type phenomenological models for the nucleon, the model parameters are fitted using the data on form factors and parton distributions.
 
Due to a multitude of complexities associated with including a gluon in a model for the nucleon, in this work, we consider a simpler, relativistic, composite spin-1/2 state, namely a quark dressed with a gluon at one loop in QCD,  using which we can perturbatively calculate the contribution of both quark and gluon OAM and spin density \cite{Harindranath:1998pc, Harindranath:1996sj}. We have recently utilized this state to study pressure and shear distributions related to gravitational form factors \cite{More:2021stk, More:2023pcy}. We compute spatial distributions on the light front using overlaps of light-front wave functions (LFWFs) within the framework of light-front Hamiltonian perturbation theory.
One key advantage is that the LFWFs for this state can be calculated analytically using the light-front QCD Hamiltonian, incorporating complete quark-gluon interactions up to one loop \cite{Harindranath:2001rc}. These LFWFs are boost-invariant since they are expressed in terms of internal coordinates, which are independent of the reference frame \cite{Brodsky:1997de}. We adopt the two-component formalism from \cite{Zhang:1993dd}, where constrained degrees of freedom are eliminated in the light-front gauge using constraint equations. This enables the analytical calculation of matrix elements for all components of the energy-momentum tensor (EMT) relevant to angular momentum distributions. Additionally, the Galilean nature of the transverse boost in light-front dynamics allows for a clear separation between the dynamics of the center of mass and the internal dynamics when calculating the longitudinal component of angular momentum \cite{Harindranath:1998ve}. In the previous work \cite{Singh:2023hgu}, we computed the spatial densities of total angular momentum coming from the quark part of the energy-momentum tensor (EMT).  We observed that these densities, when calculated using different decompositions do not agree. We also calculated the missing superpotential term in the dressed quark state which is responsible for this disparity. In this work, we present the spatial densities of various angular momentum decompositions coming from the gluon part of the EMT, and by combining the results of both papers, we verify the integrated spin sum rule for each of the decompositions.

The paper is arranged in the following manner: In Sec. \ref{sec2}, we review the different decompositions of angular momentum density in the literature; in section III, we define angular momentum densities in the front form; in section IV, we describe the two-component formalism and a dressed quark state in the light-front Hamiltonian approach; in section V, we present the calculation of angular momentum densities in the front form; in section VI, we verify the spin sum rules. Numerical results are given in section VII and the conclusion in section VIII.

\section{Energy-momentum and generalized angular momentum tensors} \label{sec2}

 The Lorentz transformation affects a multi component field in two ways: by shifting the space-time coordinates on which the fields depend and by mixing different components of fields among each other. By virtue of Noether's theorem, requiring the field theory to be invariant under this transformation leads to a conserved current, identified as the generalized angular momentum tensor ($J^{\mu \nu \rho}$). Similarly, invariance under space-time translations yields another conserved current, the energy-momentum tensor (EMT). These tensors are related to each other
 \begin{align}
    J^{\mu\nu\rho}= x^{\nu}T^{\mu\rho}-x^{\rho}T^{\mu\nu}+ S^{\mu\nu\rho} = L^{\mu\nu\rho} + S^{\mu\nu\rho}, 
\end{align}
 where $T^{\mu \nu}$, $L^{\mu \nu \rho}$ and $S^{\mu \nu \rho}$ denote the EMT, OAM and intrinsic spin angular momentum tensor respectively. In QCD, multiple decompositions of EMT are possible. This in turn leads to non-uniqueness in the way the angular momentum of a nucleon is shared among its constituents i.e., quarks and gluons.

The QCD Lagrangian can be written as,
\begin{align}
    \mathcal{L}_{\text{QCD}} = \overline{\psi} \left(\frac{i}{2} \gamma_\mu \overleftrightarrow{\partial}^\mu - m \right) \psi + g \overline{\psi}\gamma_\mu A^\mu \psi - \frac{1}{2}\text{Tr}\left[G^{\mu \nu} G_{\mu \nu}\right],
\end{align}
where  $\overleftrightarrow{\partial}^{\mu}=\overrightarrow{\partial}^{\mu}-\overleftarrow{\partial}^{\mu}$ and the field strength is $G^{\mu\nu}=\partial^{\mu}A_a^{\nu}-\partial^{\nu}A_a^{\mu}-ig\left[A_a^{\mu},A_b^{\nu}\right]$.
We begin with the Belinfante-improved EMT \cite{Belinfante1939OnTS, BELINFANTE1940449} which can be derived from this Lagrangian by considering the EMT as a conserved current under local spacetime translations \cite{Freese:2021jqs}
\begin{align}
   T^{\mu\nu}_{\text{\text{Bel}}}&=T^{\mu \nu }_{\text{Bel,q}}+T^{\mu \nu}_\text{Bel,g} \nonumber \\ &=\frac{1}{4}\overline{\psi}\left(\gamma^{\mu}i\overleftrightarrow{D}^{\nu}+\gamma^{\nu}i\overleftrightarrow{D}^{\mu}\right)\psi-2\text{Tr}\left[G^{\mu\lambda}G^{\nu}_{\lambda}\right], \label{TBel}
\end{align}
where $\overleftrightarrow{D}^\mu = \overleftrightarrow{\partial}^\mu - 2ig A^{\mu}$ for derivatives acting on quark fields. $T^{\mu \nu}_{\text{Bel}}$ is symmetric and gauge-invariant. Another rigorous way to obtain the Belinfante EMT is through functional variation of an action of QCD coupled to a weak external gravitational field with respect to the metric \cite{Polyakov:2018zvc, Blaschke:2016ohs}. Using this EMT, we can find an expression for the angular momentum density that is also symmetric and gauge-invariant
\begin{align}
    J^{\mu \nu \rho }_{\text{Bel}} &= x^\nu T^{\mu \rho}_{\text{Bel}} - x^\rho T^{\mu \nu}_{\text{Bel}} \nonumber \\
    &= \frac{1}{4} \overline{\psi} \gamma^\mu x^{[\nu}i\overleftrightarrow{D}^{\rho]} \psi+\frac{1}{4} x^{[\nu} \overline{\psi} \gamma^{\rho]} i \overleftrightarrow{D}^\mu \psi - 2\text{Tr} \left[ G^{\mu \lambda} x^{[\nu} G^{\rho ]}_\lambda \right]. \label{JBel}
\end{align}
Due to the symmetric nature of the EMT, the spin of the nucleon can only be decomposed into the total angular momentum contributions from quarks and gluons. These contributions cannot be further broken down into orbital and spin parts because the anti-symmetric component of the EMT, which is associated with spin density, is absent in the EMT
\begin{align}
    \partial_\mu J^{\mu\alpha\beta}=0 \implies \partial_\mu (L^{\mu\alpha\beta} + S^{\mu\alpha\beta}) = 0 \implies 
		\partial_\mu L^{\mu\alpha\beta} = - \partial_\mu (S^{\mu\alpha\beta}) \implies 
		T^{\alpha\beta} - T^{\beta \alpha} = - \partial_\mu (S^{\mu \alpha\beta}).
\end{align}
However, it is possible to further decompose the total angular momentum coming from the quark part into orbital and spin contribution by the subtraction of a boundary term or what are also called superpotentials. This procedure gives a new decomposition known as kinetic EMT \cite{Ji_1997}
\begin{align}
    T^{\mu \nu}_{\text{kin}} &= T^{\mu \nu}_{\text{Bel}} - \frac{1}{2} \partial_\lambda S^{\lambda \mu \nu}_{\text{q}} = T^{\mu \nu}_{\text{kin,q}} + T^{\mu \nu}_{\text{kin,g}}  \label{kin&Bel} \\
     &= \frac{1}{2} \overline{\psi} \gamma^\mu i \overleftrightarrow{D}^\nu \psi -2\text{Tr}\left[G^{\mu\lambda}G^{\nu}_{\lambda}\right],  \label{T_kin}
\end{align}
where $S^{\lambda \mu \nu }_{\text{q}} = \frac{1}{2}\epsilon^{\lambda \mu \nu \rho} \overline {\psi} \gamma_\rho \gamma_5 \psi $. Ji proposed the use of asymmetric EMT and angular momentum tensor
\begin{align}
    J^{\mu \nu \rho}_{\text{kin}} &= L^{\mu \nu \rho}_{\text{kin,q}} + S^{\mu \nu \rho}_{\text{kin,q}} + J^{\mu \nu \rho}_{\text{kin,g}} \nonumber \\
    &= \frac{1}{2} \overline{\psi} \gamma^\mu x^{[\nu} i \overleftrightarrow{D}^{\rho]} \psi + \frac{1}{2} \epsilon^{\mu \nu \rho \sigma}\overline{\psi} \gamma_\sigma \gamma_5 \psi - 2\text{Tr} \left[ G^{\mu \lambda} x^{[\nu} G^{\rho ]}_\lambda \right] . \label{J_kin}
\end{align}
Using Eq. \eqref{JBel} and \eqref{J_kin}, it can be shown that the Belinfante and kinetic decompositions are related as 
\begin{align}
    J^{\mu \nu \rho}_{\text{Bel,q}} - M^{\mu \nu \rho}_{\text{q}} &= L^{\mu \nu \rho}_{\text{kin,q}} + S^{\mu \nu \rho}_{\text{kin,q}}\,, \,\,\,\,\,\,\,\,\,\,\,\,\,\,\,\,\, J^{\mu \nu \rho}_{\text{Bel,g}} = J^{\mu \nu \rho}_{\text{kin,g}}, 
\end{align}
where $ M^{\mu \nu \rho}_{q} = \frac{1}{2}\partial_{\lambda} \left[x^\nu S^{\lambda \mu \rho}_{\text{q}} - x^\rho S^{\lambda \mu \nu}_{\text{q}}\right]$.

However, even in the kinetic decomposition, the total angular momentum coming from the gluon part of the EMT is not decomposed further. It is because of the absence of an antisymmetric part in the gluon EMT that is associated with the total derivative of the gluon spin density. This can be remedied by subtraction of a superpotential corresponding to gluon
\begin{align}
    T^{\mu \nu}_{\text{kin,g}} - 2\partial_\lambda \text{Tr} \left[ G^{\mu \lambda} A^\nu \right]
    = -2\text{Tr}\left[G^{\mu\lambda}\partial^\nu A_\lambda \right] - g \overline{\psi} \gamma^\mu A^\nu \psi ,
\end{align}
where we have used the relation $ G^{\nu}_\lambda = \partial^\nu A_\lambda - \partial_\lambda A^\nu - ig [A^\nu, A_\lambda] = \mathcal{D}^\nu A_\lambda - \partial_\lambda A^\nu $ since $ \mathcal{D}^\nu A_\lambda = \partial^\nu A_\lambda - ig [A^\nu, A_\lambda]$ for derivatives acting on gluon fields and the QCD equation of motion 
\begin{align}
    2\text{Tr} \left[ \mathcal{D}_\lambda G^{\mu \lambda} A^\nu \right] = -g \overline{\psi} \gamma^\mu A^\nu \psi.
\end{align}
Thus subtracting a superpotential from the full kinetic EMT, we get
\begin{align}
    T^{\mu \nu}_{\text{kin}} - 2\partial_\lambda \text{Tr} \left[ G^{\mu \lambda} A^\nu \right] &= \frac{1}{2} \overline{\psi} \gamma^\mu i \overleftrightarrow{\partial}^\nu \psi -2\text{Tr}\left[G^{\mu\lambda}\partial^\nu A_\lambda \right] = T^{\mu \nu}_{\text{can,q}} +  T^{\mu \nu}_{\text{can,g}}= T^{\mu \nu}_{\text{can}} \label{kin&can}
\end{align}
where $T^{\mu \nu}_{\text{can}}$ is the canonical EMT which was first proposed by Jaffe-Manohar \cite{Jaffe:1989jz}. The term $g\overline{\psi} \gamma^\mu A^\nu \psi$ gives rise to the potential angular momentum \cite{Burkardt:2012sd, Amor-Quiroz:2020qmw}.
Similar to the Belinfante EMT, the canonical EMT can also be obtained directly by using Noether's theorem but now by demanding invariance under global instead of local spacetime translations \cite{Freese:2021jqs}. 
Nonetheless, the calculation shown above illuminates the fact that different EMTs are related by superpotential terms. 
The generalized angular momentum tensor obtained from canonical EMT given in Eq. \eqref{kin&can} is 
\begin{align}
    J^{\mu \nu \rho}_{\text{can}} &= L^{\mu \nu \rho}_{\text{can,q}} + S^{\mu \nu \rho}_{\text{can,q}}  + L^{\mu \nu \rho}_{\text{can,g}}+S^{\mu \nu \rho}_{\text{can,g}}\nonumber \\ 
    &= \frac{1}{2} \overline{\psi} \gamma^\mu x^{[\nu} i \overleftrightarrow{\partial}^{\rho]} \psi + \frac{1}{2}\epsilon^{\mu\nu\rho\sigma}\overline{\psi}\gamma_{\sigma}\gamma_{5}\psi - 2\text{Tr}\left[ G^{\mu \lambda} x^{[\nu}\partial^{\rho]} A_\lambda \right] - 2\text{Tr}\left[ G^{\mu [ \nu} A^{\rho]} \right]  \label{Jcan}
\end{align}
Comparing the kinetic and canonical decompositions we get
\begin{align}
L^{\mu \nu \rho}_{\text{can,q}} + P^{\mu \nu \rho}_{\text{q}} = L^{\mu \nu \rho}_{\text{kin,q}}\,, \,\,\,\,\,\,\,\,\,\,\,\,\,\, J^{\mu \nu \rho}_{\text{kin,g}} + M^{\mu \nu \rho}_{\text{g}} &= L^{\mu \nu \rho}_{\text{can,g}} + S^{\mu \nu \rho}_{\text{can,g}}
\end{align}
where $P^{\mu \nu \rho}_{\text{q}} = g \overline{\psi} \gamma^\mu x^{[\nu}A^{\rho]} \psi$ is the potential angular momentum and $M^{\mu \nu \rho}_\text{g} = - 2\partial_\lambda \text{Tr} \left[ G^{\mu \lambda} x^{[\nu}A^{\rho]} \right]$ is the superpotential associated with gluon.

Although a complete decomposition of angular momentum is obtained using the canonical EMT, it is not gauge-invariant because it contains an ordinary derivative instead of a covariant one. To decompose the gluon part of EMT in a gauge-invariant way, one can follow the covariant formulation proposed by Wakamatsu \cite{Wakamatsu:2010cb}. It is based on the idea of decomposing gluon fields into two parts \cite{Chen:2008ag, Chen:2009mr}
\begin{align}
    A^\mu = A^\mu_{\text{pure}} + A^\mu_{\text{phys}}
\end{align}
where $A^\mu_{\text{pure}}$ is a pure-gauge term that transforms like the full gauge potential and has a null field strength
\begin{align}
    G^{\mu \nu}_{\text{pure}} = \partial^\mu A^\nu_{\text{pure}} - \partial^\nu A^\mu_{\text{pure}} - ig \left[ A^\mu _{\text{pure}}, A^\nu _{\text{pure}}\right] = \mathcal{D}^\mu_{\text{pure}}A^\nu_{\text{pure}} - \partial^{\nu}A^\mu_{\text{pure}} = 0, \label{purefieldstrength}
\end{align}
and $A^\mu _{\text{phys}}$ is the physical part that transform covariantly. The gauge transformation properties of the two parts are:
\begin{align}
    A^\mu _{\text{pure}} \longrightarrow U \left( A^\mu _{\text{pure}} + \frac{i}{g} \partial^\mu \right) U^{-1}, \,\,\,\,\,\,\,\,\,\,\,\,\,
    A^\mu _{\text{phys}} \longrightarrow U A^\mu _{\text{phys}} U^{-1}.
\end{align}
Two decompositions of EMT--gauge-invariant canonical (gic) and gauge-invariant kinetic (gik)--are obtained using this procedure each of which decomposes the nucleon spin into five components: spin and orbital AM of quarks and gluons and potential angular momentum. The Chen \textit{et. al.} or gic decomposition is a gauge-invariant extension (GIE)of the canonical decomposition and it is given as
\begin{align}
    T^{\mu \nu}_{\text{gic}} &= T^{\mu \nu}_{\text{gic,q}} + T^{\mu \nu}_{\text{gic,g}} \nonumber \\
    &=\frac{1}{2} \overline{\psi} \gamma^\mu i\overleftrightarrow{D}^\nu_{\text{pure}} \psi - 2\text{Tr} [G^{\mu\lambda}(\mathcal{D}^\nu_{\text{pure}} A_\lambda^{\text{phys}})] \label{Tgic}
\end{align}
where $D^\mu_{\text{pure}}=\partial^\mu - 2igA^\mu_{\text{pure}}$ and $\mathcal{D}^\mu_{\text{pure}}A^{\text{phys}}_\lambda =\partial^\mu A^{\text{phys}}_\lambda - 2ig[A^\mu_{\text{pure}},A^{\text{phys}}_\lambda]$. Using this EMT, we obtain the gauge-invariant canonical angular momentum tensor 
\begin{align}
    J^{\mu \nu \rho}_{\text{gic}}&=\frac{1}{2} \overline{\psi}\gamma^\mu x^{[\nu} i \overleftrightarrow{D}^{\rho]}_{\text{pure}} \psi + \frac{1}{2}\epsilon^{\mu\nu\rho\sigma}\overline{\psi}\gamma_{\sigma}\gamma_{5}\psi - 2\text{Tr}\left[ G^{\mu \lambda} x^{[\nu}\mathcal{D}^{\rho]}_{\text{pure}} A_\lambda^{\text{phys}} \right] - 2\text{Tr}\left[ G^{\mu [ \nu} A^{\rho]}_{\text{phys}} \right] \label{Jgic}
\end{align}
The canonical and gic decompositions are also related by a superpotential 
\begin{align}
    T^{\mu \nu}_{\text{gic}} &= T^{\mu \nu}_{\text{can}} + 2\partial_\lambda \text{Tr}\left[ G^{\mu \lambda} A^\nu_{\text{pure}}\right]\\
    J^{\mu \nu \rho}_{\text{gic}} &= J^{\mu \nu \rho}_{\text{can}} + 2\partial_\lambda \text{Tr}\left[ G^{\mu \lambda} x^{[\nu}A^{\rho]}_{\text{pure}}\right]
\end{align}
The gik EMT i.e., GIE of kinetic decomposition, \cite{Wakamatsu:2010cb, Wakamatsu:2010qj} can be obtained from the gic EMT just by subtracting 
 $g\overline{\psi} \gamma^\mu A^\nu_{\text{phys}} \psi$ i.e.,  the potential momentum, from the quark part and adding it to the gluon part using the QCD equation of motion. The EMT and generalized angular momentum tensor is thus given as 
\begin{align}
T^{\mu \nu}_{\text{gik}} &= \frac{1}{2} \overline{\psi} \gamma^\mu i\overleftrightarrow{D}^\nu \psi + 2\text{Tr} \left[ (\mathcal{D}_\lambda) G^{ \lambda \mu} A^\nu_{\text{phys}}\right] - 2\text{Tr} [G^{\mu\lambda}(\mathcal{D}^\nu_{\text{pure}} A_\lambda^{\text{phys}})] \\
    J^{\mu \nu \rho}_{\text{gik}} &= \frac{1}{2} \overline{\psi} \gamma^\mu x^{[\nu} i \overleftrightarrow{D}^{\rho]} \psi + \frac{1}{2}\epsilon^{\mu\nu\rho\sigma}\overline{\psi}\gamma_{\sigma}\gamma_{5}\psi  -2\text{Tr}\left[G^{\mu \lambda} x^{[\nu}\mathcal{D}^{\rho]}_{\text{pure}}A^{\text{phys}}_\lambda - (\mathcal{D}_\lambda G^{\lambda \mu})x^{[\nu}A^{\rho]}_{\text{phys}}\right] - 2\text{Tr}\left[ G^{\mu [ \nu} A^{\rho]}_{\text{phys}} \right] \label{Jgik}
\end{align}

In summary, the decompositions can be divided into two categories, kinetic and canonical \cite{Leader:2013jra}. The kinetic class includes Belinfante, Ji, and Wakamatsu decomposition. The canonical class includes Jaffe-Manohar and Chen et. al. decomposition.
The covariant way of decomposing the QCD angular momentum tensor leads to five separately gauge-invariant terms.
These terms are spin and orbital angular momentum (OAM) of quarks and gluons and potential angular momentum.
The potential angular momentum, being gauge-invariant in itself, can be added to the OAM of the quark or gluon part giving canonical and kinetic families respectively. All these decompositions differ from each other by superpotential terms at the level of densities. Thus, it is interesting to investigate the effect of the superpotentials on the distributions of angular momentum.


\section{Angular momentum distributions in light-front formalism}

To calculate the spatial distribution of angular momentum in different decompositions, we evaluate the matrix element of the angular momentum tensor. However, when such matrix elements are computed between plane wave states, the orbital angular momentum operator $x^\nu T^{\mu \rho}-x^\rho T^{\mu \nu}$ leads to ambiguities, as the spatial integration of the EMT can yield either infinite or zero values \cite{Leader:2013jra}. A standard solution is to use wave packet states \cite{Bakker:2004ib}. In this work, we start with the off-forward matrix element of the EMT following the approach of \cite{Jaffe:1989jz, Shore:1999be}; this method resolves the ambiguity and also relates the matrix element of the EMT to angular momentum densities \cite{Bakker:2004ib, Leader:2013jra, Lorce:2017wkb}. We use this approach to evaluate angular momentum distributions in light-front (LF) framework. As discussed in the Introduction,  this avoids issues related to relativistic effects in the Breit frame, which are not present in the front form due to the Galilean nature of the transverse Lorentz subgroup \cite{Brodsky:1997de}. We begin with the definition of OAM distribution 
\begin{align}
\langle L^z \rangle(x) 
=& \epsilon^{3jk}x_{\perp}^{j}\int \frac{d\Delta^+d^2\bsD}{(2\pi)^3}e^{i\Delta \cdot x}\langle T^{+k}\rangle_{\text{LF}}, \label{onshell}
\end{align}
where $\langle T^{+k}\rangle_{\text{LF}}=\dfrac{\langle p^{\prime},\boldsymbol{s}|T^{+k}(0)|p,\boldsymbol{s}\rangle}{2\sqrt{p^{\prime+}p^+}}$.

We have taken the momentum of the initial state to be $p$ and the final state $p'$. The helicities, $s$ of the initial and final states are the same as the operator structure of the angular momentum component does not involve a helicity flip. We define the average momentum of the initial and final states to be $P$ and the momentum transfer $\Delta$: \begin{align} 
p^{\mu} = \left(p^+, \boldsymbol{p}^{\perp}, p^{-}\right), \,\,\,\,\,\,\,\,\,\,\,\,\,\,\,\,\,\,\,\,\,
P^{\mu}=\frac{1}{2}\left(p^{\prime}+p\right)^{\mu}, \,\,\,\,\,\,\,\,\,\,\,\,\,\,\,\,\,\,\,\,\, \Delta^{\mu}=\left(p^{\prime}-p\right)^{\mu}.
\end{align}
The initial and final state are on-mass-shell, this gives the constraint $ P \cdot \Delta =0$ and $P^2=m^2-\dfrac{\Delta^2}{4}$. Using these relations,  we get the light-front energy transfer $\Delta^{-}$ and the average light-front energy $P^-$ as 
\begin{align}
\Delta^{-}=\frac{\bsP \cdot \bsD-P^{-}\Delta^+}{P^+}, \,\,\,\,\,\,\,\,\,\,\,\,\,
P^{-}=\frac{1}{2}\left[\frac{\left(\bsP+\frac{\bsD}{2}\right)^2+m^2}{\left(P^++\frac{\Delta^+}{2}\right)}+\frac{\left(\bsP-\frac{\bsD}{2}\right)^2+m^2}{\left(P^+-\frac{\Delta^+}{2}\right)}\right].
\end{align}
We calculate spatial densities in LF formalism using the Drell-Yan (DY) frame where $\Delta^{+}=0$ and $\bsP=\boldsymbol{0}^{\perp}$. Thus,
	\begin{align}
	    p^{\mu} = \bigg(P^+, -\frac{\bsD}{2}, \ \frac{1}{2P^+}\left(m^2+\frac{\boldsymbol{\Delta}^{\perp2}}{4}\right)\bigg),\,\,\,\,\,\,\,\,\,\,\,\,\,\,
	p^{\prime\mu}=\bigg(P^+, \frac{\bsD}{2}, \ \frac{1}{2P^+}\left(m^2+\frac{\boldsymbol{\Delta}^{\perp2}}{4}\right)\bigg),
	\end{align}
	and the invariant momentum transfer
	\be\label{momtranfer}
	\Delta^\mu\es(p^{\prime}-p)^\mu=\bigg(0, \ \bsD,   0\bigg).
	\ee
In this frame, we find that
\begin{align}
\frac{\partial}{\partial \Delta_{\perp}^j}e^{i \Delta \cdot x}=\frac{\partial}{\partial \Delta_{\perp}^j}e^{i \left(\frac{\Delta^+ x^-}{2} +\frac{\Delta^- x^+}{2} -\Delta^j_\perp x^j_\perp \right) }=-i x_{\perp}^je^{-i\Delta.x} \implies i\frac{\partial}{\partial \Delta_{\perp}^j}e^{i \Delta \cdot x}= x_{\perp}^je^{i\Delta \cdot x}, \nonumber 
\end{align}
and we obtain
\begin{align}
~~~~  \langle L^z \rangle(x)=
-i\epsilon^{3jk}\int \frac{d\Delta^+d^2\bsD}{(2\pi)^3}e^{i\Delta \cdot x} \left[\frac{\partial \langle T^{+k}\rangle_{\text{LF}} }{\partial \Delta_{\perp}^{j}}\right].
\end{align}
 Now, integrating the above expression over $x^-$ which means taking $\Delta^+ = 0$, and using Eq. \eqref{onshell}, we get
\begin{align}
\int dx^- \langle L^z \rangle(x)=\langle L^z \rangle(\boldsymbol{b}^{\perp})= -i\epsilon^{3jk}\int \frac{d^2\bsD}{(2\pi)^2}e^{-i\bsD \cdot \boldsymbol{b}^{\perp}} \left[\frac{\partial \langle T^{+k}\rangle_{\text{LF}} }{\partial \Delta_{\perp}^{j}}\right], \label{Tdist}
\end{align} 
where $\boldsymbol{b}^{\perp}$ is the impact parameter. As shown in Ref. \cite{Leader:2013jra}, zero energy transfer ensures that the matrix elements of angular momentum components of quark and gluon both are separately time-independent. Similarly, the expression of spatial distribution of intrinsic spin is given as 
\begin{align}
\langle S^z \rangle(\boldsymbol{b}^{\perp})= \frac{1}{2}\epsilon^{3jk}\int \frac{d^2\bsD}{(2\pi)^2}e^{-i\bsD \cdot \boldsymbol{b}^{\perp}} \langle S^{+jk}\rangle_{\text{LF}} . \label{Sdist}
\end{align}
For any decompositions in which the total angular momentum of a quark or gluon is not decomposed further into orbital and spin contributions, the impact-parameter distribution for total angular momentum is given as
\begin{align}
\langle J^z \rangle(\boldsymbol{b}^{\perp})= -i\epsilon^{3jk}\int \frac{d^2\bsD}{(2\pi)^2}e^{-i\bsD \cdot \boldsymbol{b}^{\perp}} \left[\frac{\partial \langle T^{+k}_{\text{Bel}}\rangle_{\text{LF}} }{\partial \Delta_{\perp}^{j}}\right]. \label{Jdist}
\end{align}

The general form of the impact-parameter distribution of the superpotential differs from the other expressions of distributions shown above. So, we derive it using the general form of total divergence terms i.e. $ M^{\mu \nu \rho }=\kappa \partial_\alpha \left(  x^\nu S^{\alpha \mu \rho} - x^\rho S^{\alpha \mu \nu}  \right)$, where $\kappa $ is a constant. This constant is just introduced because as seen from Eq. \eqref{kin&Bel} and \eqref{kin&can}, the prefactor to the total divergence term is different for quark and gluon.
To derive the distribution of the z-component of the total divergence term we have to use the component $M^{+jk}$:
\begin{align*}
    M^{z}(x) =& \frac{\kappa}{2} \epsilon^{3jk} \int \frac{d\Delta^{+}d^2 \bsD}{(2\pi)^{3}2\sqrt{p^{+}p^{\prime +}}}~ e^{i \Delta \cdot x}\partial_{l} \langle p^{\prime}, \bs{s} |x^j S^{l + k}- x^k S^{l + j}|p, \bs{s} \rangle \\
    =& - \frac{\kappa}{2} \epsilon^{3jk} \int \frac{d\Delta^{+}d^2 \bsD}{(2\pi)^{3}2\sqrt{p^{+}p^{\prime +}}}~ \left(i\Delta_{l}\right)\left[\left(i\frac{\partial ~e^{i \Delta \cdot x}}{\partial \Delta^{j}}\right) \langle p^{\prime}, \bs{s} |S^{l + k} |p, \bs{s} \rangle- \left(i\frac{\partial ~e^{i \Delta \cdot x}}{\partial \Delta^{k}}\right) \langle p^{\prime}, \bs{s} | S^{l + j}|p, \bs{s} \rangle \right]
    \\ =&  \frac{\kappa}{2}  \int \frac{d\Delta^{+}d^2 \bsD}{(2\pi)^{3}2\sqrt{p^{+}p^{\prime +}}}~ \left(\Delta_{l}\right)\left[- \epsilon^{3jk} e^{i \Delta \cdot x} \frac{\partial }{\partial \Delta^{j}} \langle p^{\prime}, \bs{s} |S^{l + k} |p, \bs{s} \rangle + \epsilon^{3kj} e^{i \Delta \cdot x} \frac{\partial ~}{\partial \Delta^{j}} \langle p^{\prime}, \bs{s} | S^{l + j}|p, \bs{s} \rangle \right]
    \\ =& \kappa \epsilon^{3jk} \int \frac{d\Delta^{+}d^2 \bsD}{(2\pi)^{3}2\sqrt{p^{+}p^{\prime +}}}~e^{i \Delta \cdot x}  \left(\Delta^{l}\right)\frac{\partial }{\partial \Delta^{j}} \langle p^{\prime}, \bs{s} |S^{l + k} |p, \bs{s} \rangle
\end{align*}
\begin{align}
   \therefore \langle M^{z}\rangle (\bsb) = \kappa \epsilon^{3jk} \int \frac{d^2 \bsD}{(2\pi)^{2}}~e^{-i \boldsymbol{\Delta}_\perp \cdot \boldsymbol{b}_\perp}  ~\Delta^{l}\frac{\partial }{\partial \Delta^{j}} \langle S^{l+k}  \rangle . \label{Mzgeneral}
   \end{align}

\section{Dressed quark state and two-component formalism}

As discussed in the introduction, to investigate the contribution to the angular momentum distributions 
coming from the gluon, instead of a nucleon state, we choose a quark dressed with a gluon. In comparison to a nucleon, it is a simple state with a gluonic degree of freedom \cite{Harindranath:1998pd, Harindranath:2001rc}. This composite state with momentum $p$ and helicity $\sigma$ is defined as 
\begin{align}
|p,\sigma \rangle = \psi_1(p,\sigma)b^{\dagger}_{\sigma}(p)|0\rangle +\sum_{\lambda_1,\lambda_2}\int \frac{dk_1^+d^2k_1^{\perp}dk_2^{+}d^2k_2^{\perp}}{(16\pi^3)\sqrt{k_1^+k_2^+}}~ \sqrt{16\pi^3 p^+}\psi_2(p,\sigma|k_1,\lambda_1;k_2,\lambda_2)\delta^{(3)}(p-k_1-k_2)b_{\lambda_1}^{\dagger}(k_1)a_{\lambda_2}^{\dagger}(k_2)|0 \rangle. 
\label{state}
\end{align}
	In Eq. \ref{state}, $\psi_1(p, \sigma)$ in the first term, corresponds to the single particle Fock space contribution with momentum (helicity) $p (\sigma)$. The two-particle LFWF, $\psi_2(p,\sigma|k_1,\lambda_1;k_2,\lambda_2)$ is related to the probability amplitude of finding two particles namely a quark and a gluon with momentum (helicity) $k_1(\lambda_1)$ and $k_2(\lambda)$,
	$b^\dagger$ and $a^\dagger$ correspond to the creation operator of quark and gluon respectively. 
	
	The LFWFs can be written in terms of relative momenta so that they are independent of the momentum of the composite state \cite{Brodsky:1997de}. This is due to the Galilean transverse boost in the light-front framework. The relative momenta $x_i$, $\bska_i$ satisfy the relation $x_1+x_2=1$ and $\bska_1+\bska_2=0$ :
	\be
	k_i^+=x_ip^+,~~~~ \bsk_i=\bska_i+x_i \bsp, 
	\ee 
	where $x_i$ is the longitudinal momentum fraction for the quark or gluon, inside the two-particle LFWF.
	The boost invariant two-particle LFWF can be written as,
	\begin{align}
		\phi_{\lambda_1,\lambda_2}^{\sigma}(x,\bska)=& \frac{g}{\sqrt{2(2\pi)^3}}\bigg[\frac{x(1-x)}{\kappa_{\perp}^2+m^2(1-x)^2}\bigg]\frac{T^a}{\sqrt{1-x}} \nonumber\\  
  & \chi_{\sigma_1}^{\dagger}\bigg[-\frac{2(\bska\cdot \boldsymbol{\epsilon}_{\lambda_2}^{\perp*})}{1-x}-\frac{1}{x}(\tilde{\sigma}^{\perp}\cdot\bska)(\tilde{\sigma}^{\perp}\cdot \boldsymbol{\epsilon}_{\lambda_2}^{\perp*})+im(\tilde{\sigma}^{\perp}\cdot \boldsymbol{\epsilon}_{\lambda_2}^{\perp*})\frac{1-x}{x}\bigg]\chi_{\sigma} \psi_1^{\sigma}, \label{qdress}  
	\end{align}
	where, $\phi^{\sigma }_{\lambda_1,\lambda_2}(x_i,\bska_i)=\sqrt{P^+}\psi_2 (P,\sigma|k_1,\lambda_1;k_2,\lambda_2)$,  $g$ is the quark-gluon  coupling. $T^a$ and $\boldsymbol{ \epsilon}_{\lambda_2}^\perp$ are colour SU(3) matrices and polarization vector of the gluon. The quark mass and the two-component spinor for the quark are denoted by $m$ and $\chi_\lambda$ respectively, $\lambda=1,2$ correspond to helicity up/down of the quark. We have used the notation $\tilde{\sigma}_1=\sigma_2$ and $\tilde{\sigma}_2=-\sigma_1$ \cite{Harindranath:2001rc}. In the expression above, $x$ and $\bska$ are longitudinal momentum fraction and the relative transverse momentum of quark respectively. So, to do calculations of angular momentum of the gluon part, we substitute $x\rightarrow 1-x \text{ and } \kappa\rightarrow -\kappa$ in the Eq. \eqref{qdress} in accordance with the constraints on the relative momenta \cite{Harindranath:1998ve}
\begin{align}
		\phi_{\lambda_1,\lambda_2}^{\sigma}(x,\bska)=& \frac{g}{\sqrt{2(2\pi)^3}}\bigg[\frac{x(1-x)}{\kappa_{\perp}^2+m^2 x^2}\bigg]\frac{T^a}{\sqrt{x}} \nonumber\\  
  & \chi_{\sigma_1}^{\dagger}\bigg[\frac{2(\bska\cdot \boldsymbol{\epsilon}_{\lambda_2}^{\perp*})}{x}+\frac{1}{1-x}(\tilde{\sigma}^{\perp}\cdot\bska)(\tilde{\sigma}^{\perp}\cdot \boldsymbol{\epsilon}_{\lambda_2}^{\perp*})+im(\tilde{\sigma}^{\perp}\cdot \boldsymbol{\epsilon}_{\lambda_2}^{\perp*})\frac{x}{1-x}\bigg]\chi_{\sigma} \psi_1^{\sigma}, \label{gdress}  
\end{align}
where $\psi_1$ gives the normalization of the state
 \begin{align}
    {\mid \psi_1 \mid }^2  = 1-\frac{g^2}{8 \pi^2}C_f \int dx \left\{\frac{1+x^2}{1-x} \log{\left(\frac{\Lambda^2}{m^2 (1-x)^2}\right)} -\frac{2x}{1-x}\right\}. \label{normalization}
 \end{align}
 In the above expression, $\Lambda$ is the cutoff on the transverse momentum integration. We use the two-component formalism \cite{Zhang:1993dd}  of light-front Hamiltonian QCD, where in the light-front gauge $A^+=0$, one can eliminate the unphysical degrees of freedom using the equations of constraint. Using the light-front representation of gamma matrices, the quark field can be decomposed as $\psi=\psi^{+}+\psi^{-}$, $\psi_{\pm}=\Lambda_{\pm}\psi=\frac{1}{2}\gamma^{0}\gamma^{\pm}\psi$
\begin{align}
\psi_+= \begin{bmatrix}
\xi\\0
\end{bmatrix}, ~~~~\psi_-=\begin{bmatrix}
0\\ \eta
\end{bmatrix}, 
\end{align}
where $\Lambda_{\pm}$ are projection operator, $\xi$ represents the two-component light-front quark field and $ \eta$ is the constrained field: 
\begin{align}
	\xi(y) &= \sum_{\lambda}\chi_{\lambda}\int \frac{[dk]}{\sqrt{2(2\pi)^3}}[b_{\lambda}(k)e^{-ik\cdot y}+d^{\dagger}_{-\lambda}(k)e^{ik\cdot y}], \\
	\eta(y) &= \left(\frac{1}{i\partial^+}\right)\left[\sigma^{\perp}\cdot\left(i\partial^{\perp}+g A^{\perp}(y)\right)+im\right]\xi(y),
	\end{align}
 where $[dk]= \frac{dk^+ d^2k^\perp} {\sqrt{2(2\pi)^3{k^{+}}}}$.
 The dynamical components of the gluon field are given by 
\begin{align}
A^{\perp}(y) = \sum_{\lambda} \int \frac{[dk]}{\sqrt{2 (2 \pi)^3 k^+}}[{\bf\epsilon}^{\perp}_{\lambda}a_{\lambda}(k)e^{-i k \cdot y}+ {\bf \epsilon}^{\perp*}_{\lambda}a^{\dagger}_{\lambda}(k)e^{i k \cdot y}],
\end{align} 
where $\chi_{\lambda}$ is the eigenstate of $\sigma^3$ and $\epsilon_{\lambda}^{i}$ are the polarization vectors of the transverse gauge field. The $\psi^{-}$ component and the longitudinal component of the gauge field $A^{-}$ are constrained fields and can be written in terms of $\psi^{+}$ and $A^{\perp}$ in the following way 
\begin{align}
    i\partial^+\psi_-=&\left(i\alpha^{\perp}\cdot \partial^{\perp}+g \alpha^{\perp}\cdot A^{\perp}+\beta m\right)\psi_+,\\
    \frac{1}{2}\partial^+E_a^-=&\left(\partial^iE_a^i+gf^{abc}A_b^iE_c^i\right)-g\psi_+^{\dagger}T^a\psi_+,
\end{align}
where $T^{a}$ the Gell-Mann SU(3) matrices: $[T^{a},T^{b}]=if^{abc}T^{c}$ and $\text{Tr}(T^{a}T^{b})=\frac{1}{2}\delta_{ab}$, $m$ is the quark mass, $\alpha^{\perp}=\gamma^0\gamma^{\perp}$, $\beta=\gamma^0$ and $E_a^{-,i}=-\frac{1}{2}\partial^{+}A_a^{-,i}$, $(i=1,2)$.

Following the approach of \cite{Diehl:2002he, Chakrabarti:2005zm, More:2021stk, More:2023pcy}, to make smooth plots of the spatial distributions, we use a Gaussian wave packet state in momentum space centered at the origin. The state is confined in transverse momentum space with definite longitudinal momentum and can be expressed as
\be
\frac{1}{16\pi^3}\int \frac{d^2 \bsp dp^+}{p^+}\phi\left( p\right) \mid p^+,\bsp,\lambda \rangle 
\ee
with $\phi(p)=p^+\ \delta(p^+-p_0^+)\ \phi\left(\bsp\right)$.
We a choose a Gaussian shape for $\phi\left( \bsp\right)$ in transverse momentum :
\be\label{gaussian}
\phi\left(\bsp\right)
= e^{-\frac{\bspp}{2\sigma^2}}
\ee 
where $\sigma$ is the width of Gaussian.
\section{Angular momentum distributions from gluon part of EMT}

In this section, we discuss the spatial densities of the gluon angular momentum in a dressed quark state, by taking different decompositions. The impact-parameter distribution of gluon OAM in canonical decomposition is given by Eq. \eqref{Tdist}
\begin{align}
    \langle L^z_{\text{can,g}} \rangle(\boldsymbol{b}^{\perp})&= -i\epsilon^{3jk}\int \frac{d^2\bsD}{(2\pi)^2}e^{-i\bsD \cdot \boldsymbol{b}^{\perp}} \left[\frac{\partial \langle T^{+k}_{\text{can,g}}\rangle_{\text{LF}} }{\partial \Delta_{\perp}^{j}}\right],  \nonumber 
    \\&=  i\int \frac{d^2\bsD}{(2\pi)^2}e^{-i\bsD \cdot \boldsymbol{b}^{\perp}} \left[\frac{\partial \langle T^{+1}_{\text{can,g}}\rangle_{\text{LF}} }{\partial \Delta_{\perp}^{(2)}}-\frac{\partial \langle T^{+2}_{\text{can,g}}\rangle_{\text{LF}} }{\partial \Delta_{\perp}^{(1)}}\right]_{\text{DY}} \label{Lz expr}
\end{align} 
where $\boldsymbol{b}^{\perp}$ is the Fourier conjugate of the transverse momentum transfer $\boldsymbol{\Delta_\perp}$.
In order to calculate these densities, we must first evaluate the matrix element of the EMT to which they are associated. So, we consider the gluon part of the canonical EMT given by the Eq. \eqref{kin&can}
\begin{align}
    T^{+k}_{\text{can,g}}&=-2\text{Tr}\left[ G^{+ \alpha } \partial^k A_\alpha \right]=G_{a}^{+\perp}~\partial^{k}A_{a}^{\perp}= \left(\partial^{+}A_{a}^{\perp}\right)\left(\partial^{k}A_{a}^{\perp}\right) \\
&= \sum_{\lambda, \lambda^{\prime}}\int \frac{dk^{+}d^2\bsk dk^{+}d^2\bskpr}{\left(16\pi^3\right)^2k^{\prime +}}k^{\prime k}\bigg[-\left(\epsilon^{\perp}_{\lambda}\cdot \epsilon^{\perp}_{\lambda^{\prime}}\right)a_{\lambda}(k)a_{\lambda^{\prime}}(k^{\prime})e^{-i\left(k+k^{\prime}\right)\cdot y} 
+ \left(\epsilon^{\perp}_{\lambda}\cdot \epsilon^{\perp*}_{\lambda^{\prime}}\right)a_{\lambda}(k)a^{\dagger}_{\lambda^{\prime}}(k^{\prime})e^{-i\left(k-k^{\prime}\right)\cdot y}  \nonumber \\  & 
\hspace{5.1cm}+ \left(\epsilon^{\perp*}_{\lambda}\cdot \epsilon^{\perp}_{\lambda^{\prime}}\right)a^{\dagger}_{\lambda}(k)a_{\lambda^{\prime}}(k^{\prime})e^{i\left(k-k^{\prime}\right)\cdot y} 
 -\left(\epsilon^{\perp*}_{\lambda}\cdot \epsilon^{\perp*}_{\lambda^{\prime}}\right)a^{\dagger}_{\lambda}(k)a^{\dagger}_{\lambda^{\prime}}(k^{\prime})e^{i\left(k^{\prime}+k\right)\cdot y}\bigg]
\end{align}
We calculate the matrix elements of $T^{+k}_{\text{can,g}}(0)$ by sandwiching the above expression between dressed quark state given in Eq. \eqref{state}. Only the two-particle diagonal term is non-zero which is then expressed in terms of boost-invariant LFWF and Jacobi momenta ($x,\kappa^\perp$).  
We have shown the form of the diagonal matrix element in Eq. \eqref{T+k_can1} and \eqref{T+k_can2} in the appendix, which are obtained after performing the integration over the Jacobi momenta. Substituting these expressions in Eq.\eqref{Lz expr}, we get
\begin{align}
    &\langle L_{\text{can,g}}^z \rangle(\boldsymbol{b}^{\perp}) \nonumber \\
    &= g^2 C_f \int \frac{d^2\boldsymbol{\Delta}^{\perp}}{(2\pi)^2}e^{-i\bsb \cdot \bsD} \int \frac{dx}{8 \pi^2} \frac{x-2}{(x-1){\Delta}^2 \omega'} \left[ (2 m^2 x^2 + (x-1)^2 \Delta^2) \text{log} \left( \frac{1 + \omega'}{-1 + \omega'} \right) - (x-1)^2 \Delta^2 \omega' \text{log} \left( \frac{\Lambda^2}{m^2 x^2} \right)
 \right], \label{OAM_can}
\end{align}
where $\omega'=\sqrt{1+\frac{4m^{2}x^{2}}{\left(1-x\right)^{2}\Delta^{2}}}$ and $\Lambda$ is the ultraviolet cutoff on the transverse momentum. $\Lambda$ gives the scale dependence of the results. 

For the spatial distribution of the intrinsic part of the gluon, we write the operator structure of the gluon spin term in Eq.\eqref{Jcan} as
\begin{align}
   \nonumber S^{+jk}_{\text{can,g}}=&- 2\text{Tr}\left[ G^{+ [ j} A^{k]} \right]=-G^{+j}_{a}A^{k}_{a}+G^{+k}_{a}A^{j}_{a}=-\left(\partial^{+}A_{a}^{j}\right)A_{a}^{k}+\left(\partial^{+}A_{a}^{k}\right)A_{a}^{j}\\ \nonumber =&i\sum_{\lambda,\lambda^{\prime}}\int \frac{dk^{+}d^2\bs{k}^{\perp} dk^{\prime +}d^2\bs{k}^{\perp}}{\left(16\pi^3\right)^{2}k^{\prime +}}\bigg[\left(\epsilon^{j}_{\lambda}\epsilon^{k}_{\lambda^{\prime}}-\epsilon^{k}_{\lambda}\epsilon^{j}_{\lambda^{\prime}}\right)a_{\lambda}(k)a_{\lambda^{\prime}}(k^{\prime})e^{-i\left(k+k^{\prime}\right)\cdot y}+\left(\epsilon^{j}_{\lambda}\epsilon^{k*}_{\lambda^{\prime}}-\epsilon^{k}_{\lambda}\epsilon^{j*}_{\lambda^{\prime}}\right)a_{\lambda}(k)a^{\dagger}_{\lambda^{\prime}}(k^{\prime})e^{-i\left(k-k^{\prime}\right)\cdot y}\\ & \hspace{2cm} -\left(\epsilon^{j*}_{\lambda}\epsilon^{k}_{\lambda^{\prime}}-\epsilon^{k*}_{\lambda}\epsilon^{j}_{\lambda^{\prime}}\right)a^{\dagger}_{\lambda}(k)a_{\lambda^{\prime}}(k^{\prime})e^{i\left(k-k^{\prime}\right)\cdot y}-\left(\epsilon^{j*}_{\lambda}\epsilon^{k*}_{\lambda^{\prime}}-\epsilon^{k*}_{\lambda}\epsilon^{j*}_{\lambda^{\prime}}\right)a^{\dagger}_{\lambda}(k)a^{\dagger}_{\lambda^{\prime}}(k^{\prime})e^{i\left(k+k^{\prime}\right)\cdot y}\bigg].
\end{align}
Again, after evaluating the diagonal and off-diagonal (vanishes) matrix element in terms of LFWF and Jacobi momenta and substituting them in Eq. \eqref{Sdist}, we get
\begin{align}
  \langle S_{\text{can,g}}^z \rangle(\boldsymbol{b}^{\perp})
&= g^2 C_f \int \frac{d^2\boldsymbol{\Delta}^{\perp}}{(2\pi)^2}e^{-i\bsb \cdot \bsD} \int \frac{dx}{{8 \pi^2 \Delta^4}} \times \nonumber  \\ &\left[ \frac{{((x-2 ) (x-1)^2 \Delta^4 + 2 m^2 x^2 ((3x-4) \Delta^2 - (x-2) \Delta^2)) \log\left(\frac{{1 + \omega'}}{{-1 + \omega'}}\right)}}{{(x-1)^2 \omega'}} - (x-2) \Delta^4 \log\left(\frac{{\Lambda^2}}{{m^2 x^2}}\right) \right]. \label{spin_can}
\end{align}
Similarly, the impact-parameter distribution for the gluon part of the kinetic total angular momentum is calculated using Eq.\eqref{T_kin} and \eqref{Jdist}
\begin{align}
		\langle J^z_{\text{kin,g}} \rangle (\bsb)&= g^2 C_f \int \frac{d^2\boldsymbol{\Delta}^{\perp}}{(2\pi)^2}e^{-i\bsb \cdot \bsD} \int \frac{dx}{16 \pi^2} \frac{1}{\left(1-x\right)^2{\Delta}^4 \omega'^3} ~\times \nonumber \\
&\left[x \left(8 m^4 x^3 + 6 m^2 (x-2) x^2 \Delta^2 + (x-2) (x-1)^2 \Delta^4\right) \log\left(\frac{1 + \omega'}{-1 + \omega'}\right)\right. \nonumber \\
&\left.+ \Delta^2 \omega' \left((x-1) \left(4 m^2 x^2 + (x-2) (x-1 ) \Delta^2\right) - (x-2) x \left(4 m^2 x^2 + (x-1)^2 \Delta^2\right) \log\left(\frac{\Lambda^2}{m^2 x^2}\right)\right) \right] \label{Jzkin}
	\end{align}
The relevant $\kappa^\perp$ integrations and the steps of the calculation are given in Appen. \ref{appB}, \ref{appC} and \ref{appE}.

In order to make a proper comparison between the canonical and kinetic decomposition, it is important to include the correction term to the kinetic total angular momentum. 
This term corresponds to the superpotential given in Eq. \eqref{kin&can}. It's expression at the distribution level is given as
\begin{align}
    \langle M^{z}_{\text{g}} \rangle (\bsb) \nonumber &= \kappa \epsilon^{3jk} \int \frac{d^2 \bsD}{(2\pi)^{2}}~e^{i \boldsymbol{\Delta}_\perp \cdot \boldsymbol{b}_\perp}  ~\Delta^{l}\frac{\partial }{\partial \Delta^{j}} \langle S^{l+k}_{\text{g}}  \rangle .  \label{Mz expr}
\end{align}
Substituting the matrix element of $S^{l+k}_{\text{g}}$ given in Eq. \eqref{S1+2} and \eqref{S2+1} in the above expression, we get the spatial distribution of the superpotential term
\begin{align}
    \langle M^{z}_{\text{g}}\rangle (\bsb) =& g^2 C_f \int \frac{d^2\boldsymbol{\Delta}^{\perp}}{(2\pi)^2}e^{-i\bsb \cdot \bsD} \int \frac{dx}{8\pi^2 (1-x)^3\Delta^4 \omega'^3} \times \nonumber \\
    &\left[2 m^2 x^3 (4 m^2 x + (x - 1) \Delta^2) \log\left(\frac{{1 + \omega'}}{{-1 + \omega'}}\right) -  \Delta^2 \omega'(x - 1)^2 (4 m^2 x^2 + (x - 2) (x - 1) \Delta^2)
\right].
\end{align}
Since the gluon part of Belinfante decomposition is exactly same as that of kinetic decomposition
\begin{align}
    \langle J^z_{\text{Bel,g}} \rangle (\bsb) = \langle J^z_{\text{kin,g}} \rangle (\bsb)
\end{align}
Now consider the relevant light-front components of gic decomposition given in Eq. \eqref{Tgic} and \eqref{Jgic} 
\begin{align}
    T^{+k}_{\text{gic}} &= \frac{1}{2} \overline{\psi}  \gamma^+ i\overleftrightarrow{D}^k_{\text{pure}} \psi   - 2\text{Tr} [G^{+\lambda}(\mathcal{D}^k_{\text{pure}} A_\lambda^{\text{phys}})], \hspace{2.7cm}
    \label{T_gic} \\
    S^{+jk}_{\text{gic}}&=\frac{1}{2}\epsilon^{+jk-}\overline{\psi}\gamma_{-}\gamma_{5}\psi - 2\text{Tr}\left[ G^{+ [ j} A^{k]}_{\text{phys}} \right],
    \label{J_gic}
\end{align}  
and the relevant light-front components of gik decomposition given in Eq. \eqref{Jgik}
\begin{align}
    T^{+k }_{\text{gik}} &= \frac{1}{2} \overline{\psi}  \gamma^+ i\overleftrightarrow{D}^k \psi + 2\text{Tr} \left[ (\mathcal{D}_\lambda G^{ \lambda +}) A^k_{\text{phys}}\right] - 2\text{Tr} [G^{+\lambda}(\mathcal{D}^k_{\text{pure}} A_\lambda^{\text{phys}})],
    \label{T_gik} \\
    S^{+jk}_{\text{gik}} &=  \frac{1}{2}\epsilon^{+jk-}\overline{\psi}\gamma_{-}\gamma_{5}\psi - 2\text{Tr}\left[ G^{+ [ j} A^{k]}_{\text{phys}} \right],
    \label{J_gik}
\end{align}
where $A_{\text{\text{pure}}}^{k}=A^{k}-A_{\text{\text{phys}}}^{k}$. For the transverse component of the gauge field $A^{k}=A_{\text{\text{phys}}}^{k}$. So in light-front, the gic decomposition of EMT is the same as the canonical one. 
Thus, the OAM and spin distribution of quark and gluon calculated from the gic decomposition of EMT is the same as the canonical OAM and spin distribution. Due to the same reason, distributions in gik decomposition coincide with the kinetic one in the light-front gauge, for a dressed quark.

\section{Verification of the spin sum rule} 

 In the section, we verify the longitudinal spin sum rule for a dressed quark state, using different decompositions of angular momentum. The spin sum rule states that at the integrated level, the sum of all the angular momentum contributions of any decomposition should be equal to 1/2. The total angular momentum of the nucleon is related to a conserved operator, and this is independent of the renormalization scale.  However,  individually the orbital and intrinsic parts of the quark and gluon angular momentum in different decompositions are not related to conserved operators \cite{Ji:1995cu}  and their matrix elements depend on the renormalization scale. As discussed in the introduction, it is difficult to verify the spin sum rule in phenomenological spectator-type models of the nucleon. A field theory based perturbative model like ours, on the other hand, is a very useful tool to explicitly verify the sum rule at the level of one loop.
 
 \subsection{Kinetic decomposition}Mathematically, the spin sum rule for the kinetic decomposition can be given as follows:
\begin{align}
    \int d^2 \boldsymbol{b}^\perp \left[\langle L^z_{\text{kin,q}} \rangle (\bsb) + \langle S^z_{\text{kin,q}} \rangle (\bsb) + \langle J^z_{\text{kin,g}} \rangle (\bsb) \right] = \frac{1}{2}, \label{kinssr}
\end{align}
From the Eq. \eqref{Lzkin}, \eqref{Szkin} and \eqref{Jzkin}, we see that the expressions on the LHS involve a two-dimensional inverse Fourier transform over $\boldsymbol{\Delta}^\perp$. These expressions when used in the above equation lead to a simplification. Since, $\boldsymbol{b}^\perp$ is a Fourier conjugate of $\boldsymbol{\Delta}^\perp$, integrating over $\boldsymbol{b}^\perp$ amounts to taking the limit $\boldsymbol{\Delta^\perp}=0$ in these expressions. Thus we get 
\begin{align}
    \int d^2 \bsb \langle L^z_{\text{kin,q}} \rangle (\bsb) &= -\frac{g^2C_f}{8\pi^2}\left[\frac{1}{9} +\frac{2}{3}\log \left(\frac{\Lambda^2}{m^2}\right)\right], \nonumber \\
    \int d^2 \bsb \langle S^z_{\text{kin,q}} \rangle (\bsb) &= \frac{1}{2}\int dx \left[ 1-\frac{g^2 C_f}{8 \pi^2} \left\{\frac{1+x^2}{1-x} \ln{\left(\frac{\Lambda^2}{m^2 (1-x)^2}\right)} -\frac{2x}{1-x}\right\}\right. \nonumber \\
		&\hspace{1.1cm} \left.-\frac{g^2 C_f}{8\pi^2}\left\{ \frac{2(1-x+x^2)}{1-x} - \frac{1+x^2}{1-x} \log{\left( \frac{\Lambda^2}{m^2(1-x)^2} \right)} \right\} \right] \nonumber \\ &=\frac{1}{2} +  \frac{g^2 C_f}{8\pi^2}\int dx(x-1) \nonumber \\  &= \frac{1}{2} - \frac{g^2 C_f}{16\pi^2}, \nonumber \\
    \int d^2 \bsb \langle J^z_{\text{kin,g}} \rangle (\bsb) &= -\frac{g^2C_f}{8\pi^2}\int dx (1-x)\left[x-(x+1)\log\left(\frac{\Lambda^2}{m^2}\right) +2(1+x)\log(1-x) \right] \nonumber \\ &= \frac{g^2 C_f}{8\pi^2}\left[\frac{11}{18} + \frac{2}{3}\log\left(\frac{\Lambda^2}{m^2}\right)\right], \nonumber 
\end{align}
Here, $x$ is the fraction of longitudinal momentum taken by a quark, and $1-x$ is the momentum fraction of the gluon. In the expression of $\langle S^z_{\text{kin,q}} \rangle$, the first line (second line) of the integrand represents the single-particle contribution (two-particle contribution) to the matrix element. The single-particle contribution as shown in Eq. \eqref{normalization} is the normalization factor of the dressed quark state. It is crucial for verifying the spin sum rule. As seen above, the $\Lambda$-dependent term in the two-particle contribution to the quark spin distribution is precisely canceled by a $\Lambda$-dependent term coming from the single particle contribution, yielding a cutoff-independent result for the quark spin distribution.   Substituting these expressions in Eq. \eqref{kinssr}, we find that the longitudinal spin sum rule is indeed verified for the kinetic (Ji) decomposition
\begin{align}
     \langle L^z_{\text{kin,q}} \rangle + \langle S^z_{\text{kin,q}} \rangle  + \langle J^z_{\text{kin,g}} \rangle  = \frac{1}{2}.
\end{align}
In fact, by adding all the terms, all terms of $O(g^2)$ get canceled. As stated before, the total angular momentum of the dressed quark does not depend on the renormalization scale. In our approach, $\Lambda$ introduces the scale dependence. Cancellation of $\Lambda$ dependent terms verifies the scale independence of the total angular momentum. As we show below, this is true for all decompositions.

\subsection{Belinfante decomposition}
The longitudinal spin sum rule for the Belinfante decomposition is given as
\begin{align}
    \int d^2 \boldsymbol{b}^\perp \left[\langle J^z_{\text{Bel,q}} \rangle (\bsb) + \langle J^z_{\text{Bel,g}} \rangle (\bsb) \right] = \frac{1}{2}.
\end{align}
Integrating Eq. \eqref{JzBelq} and \eqref{Jzkin} over the impact parameter space gives 
\begin{align}
\int d^2 \bsb \langle J^z_{\text{Bel,q}} \rangle (\bsb) &=\frac{1}{2}\int dx \,\, \left\{1-\frac{g^2 C_f}{8 \pi^2} \left[\frac{1+x^2}{1-x} \log{\left(\frac{\Lambda^2}{m^2 (1-x)^2}\right)} -\frac{2x}{1-x}\right] \right\} \nonumber \\ & \hspace{0.4cm} -\int dx \,\, \frac{g^2C_f}{8\pi^2}\left\{ \frac{x^2(2-x)}{1-x} + \frac{(1-2x-3x^2+2x^3)}{2(1-x)}\log{\left(\frac{\Lambda^2}{m^2 (1-x)^2}\right)} \right\} \nonumber \\
&=\frac{1}{2} + \frac{g^2 C_f}{8\pi^2}\int dx \left[ x(1-x)+ (x^2-1)\log \left( \frac{\Lambda^2}{m^2(1-x)^2}\right)\right] \nonumber \\
\int d^2 \bsb \langle J^z_{\text{Bel,g}} \rangle (\bsb) &= -\frac{g^2 C_f}{8\pi^2}\int dx \left[ x(1-x)+ (x^2-1)\log \left( \frac{\Lambda^2}{m^2(1-x)^2}\right)\right].\nonumber 
\end{align}
In the expression of $ \langle J^z_{\text{Bel,q}} \rangle $,  the first line (second line) of the integrand represents the single-particle contribution (two-particle contribution) to the matrix element. 
It can be seen from above that $O(g^2)$ terms in the matrix element of quark and gluon total angular momentum exactly cancel each other thus giving the Belinfante spin sum rule
\begin{align}
    \langle J^z_{\text{Bel,q}} \rangle+ \langle J^z_{\text{Bel,g}} \rangle = \frac{1}{2}.
\end{align}
The explicit factor of $\frac{1}{2}$ is present due to the single-particle contribution to the quark total angular momentum matrix element. Thus the normalization of the dressed quark state i.e. $\psi_1$ plays a vital role in the verification of the spin sum rule.

\subsection{Canonical decomposition}
The longitudinal spin sum rule for the canonical (Jaffe-Manohar)  decomposition has the following form
\begin{align}
    \int d^2 \boldsymbol{b}^\perp \left[\langle L^z_{\text{can,q}} \rangle (\bsb) + \langle S^z_{\text{can,q}} \rangle (\bsb) + \langle L^z_{\text{can,g}} \rangle (\bsb) + \langle S^z_{\text{can,g}} \rangle (\bsb) \right] = \frac{1}{2}, \label{canssr}
\end{align}
where,
\begin{align}
    \int d^2 \bsb \langle L^z_{\text{can,q}} \rangle (\bsb) &=-\frac{g^2 C_f}{8\pi^2}\left[\frac{1}{9} +\frac{2}{3}\log \left(\frac{\Lambda^2}{m^2}\right)\right], \nonumber \\
    \int d^2 \bsb \langle S^z_{\text{can,q}} \rangle (\bsb) &= \frac{1}{2}\int dx \left[ 1-\frac{g^2 C_f}{8 \pi^2} \left\{\frac{1+x^2}{1-x} \ln{\left(\frac{\Lambda^2}{m^2 (1-x)^2}\right)} -\frac{2x}{1-x}\right\}\right. \nonumber \\
		&\hspace{1.1cm} \left.-\frac{g^2 C_f}{8\pi^2}\left\{ \frac{2(1-x+x^2)}{1-x} - \frac{1+x^2}{1-x} \log{\left( \frac{\Lambda^2}{m^2(1-x)^2} \right)} \right\} \right] \nonumber \\ &=\frac{1}{2} +  \frac{g^2 C_f}{8\pi^2}\int dx(x-1) \nonumber \\  &= \frac{1}{2} - \frac{g^2 C_f}{16\pi^2}, \nonumber \\
    \int d^2 \bsb \langle L^z_{\text{can,g}} \rangle (\bsb) &=
    \frac{g^2 C_f}{8\pi^2}\int dx \left\{x(x+1)\left[1-\log \left(\frac{\Lambda^2}{m^2}\right)+2\log\left(1-x\right)\right] \right\}\nonumber \\ 
    &= -\frac{g^2 C_f}{8\pi^2}\left[\frac{17}{9}+\frac{5}{6} \log \left(\frac{\Lambda^2}{m^2}\right) \right],\nonumber \\
    \int d^2 \bsb \langle S^z_{\text{can,g}} \rangle (\bsb) &=  -\frac{g^2 C_f}{8\pi^2}\int dx \left\{2x-(x+1)\log\left(\frac{\Lambda^2}{m^2}\right) +2(x+1)\log(1-x) \right\} \nonumber \\ 
    &= \frac{g^2 C_f}{8\pi^2}\left[\frac{5}{2}+ \frac{3}{2} \log \left(\frac{\Lambda^2}{m^2}\right)\right].\nonumber
\end{align}
Once again the single-particle contribution to the quark spin proves integral in making the quark spin independent of the cutoff. Other $\Lambda$-dependent terms in OAM of quark and OAM and spin of gluon cancel amongst each other. Substituting these components in Eq. \eqref{canssr}, we find that the canonical decomposition also follows the spin sum rule
\begin{align}
    \langle L^z_{\text{can,q}} \rangle + \langle S^z_{\text{can,q}} \rangle + \langle L^z_{\text{can,g}} \rangle + \langle S^z_{\text{can,g}} \rangle = \frac{1}{2}.
\end{align}
It is noteworthy that the $\Lambda$-dependent terms in $\langle L^z_{\text{can,q}} \rangle$, $\langle L^z_{\text{can,g}} \rangle$ and $\langle S^z_{\text{can,g}} \rangle$ align precisely with the results obtained for the canonical decomposition in \cite{Harindranath:1998ve}. However, the finite parts differ because in \cite{Harindranath:1998ve},  quark mass effects were ignored to avoid introducing higher twist contributions to the quark and gluon OAM. Another key difference is that their approach derived the canonical decomposition by using a Belinfante EMT and neglecting surface terms. While this leads to a similar result to ours when considering the total (integrated) values, as we will demonstrate in the next section, choosing the Belinfante or canonical EMT yields different results for the total angular momentum at the density level.

\section{Numerical analysis}

\begin{figure}[ht]
\begin{minipage}{0.49\linewidth}
\includegraphics[scale=0.95]{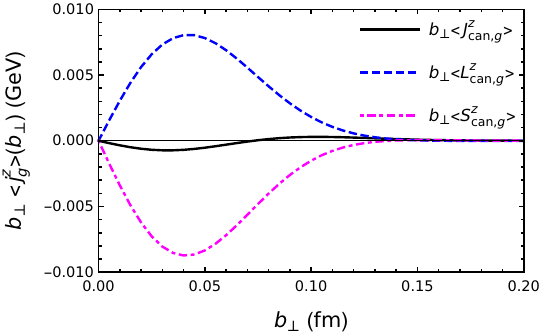}
\end{minipage}
\begin{minipage}{0.49\linewidth}
\includegraphics[scale=0.95]{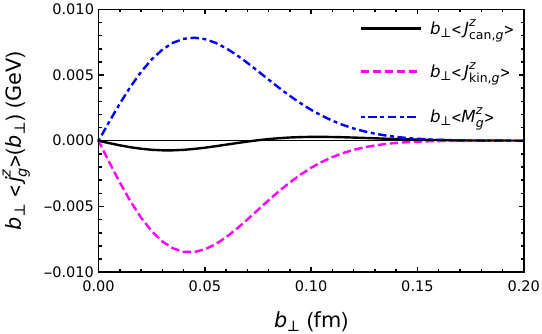}
\end{minipage}  
\caption{Plots of longitudinal angular momentum distribution of gluons as a function of $b_\perp$. Left: Sum of the $\langle L^z_{\text{can,g}} \rangle$ (dashed line) and $\langle S^z_{\text{can,g}}\rangle$ (dot-dashed line) given by $\langle J^z_{\text{can,g}} \rangle$ (solid line). Right: $\langle J^z_{\text{can,g}} \rangle$ (solid line) is given by the sum of $\langle J^z_{\text{kin,g}} \rangle$ (dashed line) and $\langle M^z_{\text{g}} \rangle$ (dot-dashed line). Here, $m = 0.3$ GeV, $g=1$, $C_f=1$, and $\Lambda = 2.63$ GeV.}
\label{gDist_graph}
\end{figure}

\begin{figure}[ht]
    \centering

    \begin{minipage}{0.45\textwidth}
        \centering
        \includegraphics[scale=0.9]{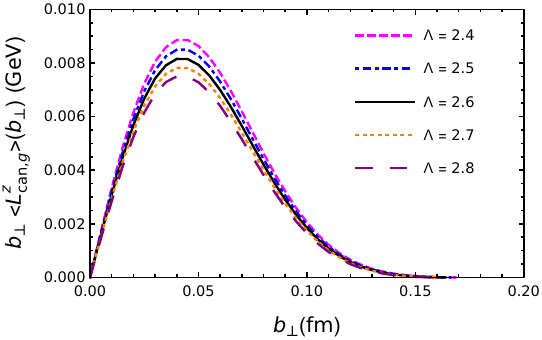}
    \end{minipage}
    \hfill
    \begin{minipage}{0.45\textwidth}
        \centering
        \includegraphics[scale=0.9]{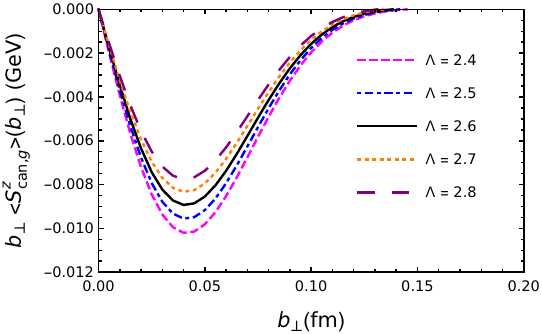}
    \end{minipage}

    \medskip 

    \begin{minipage}{\textwidth}
        \centering
        \includegraphics[scale=0.9]{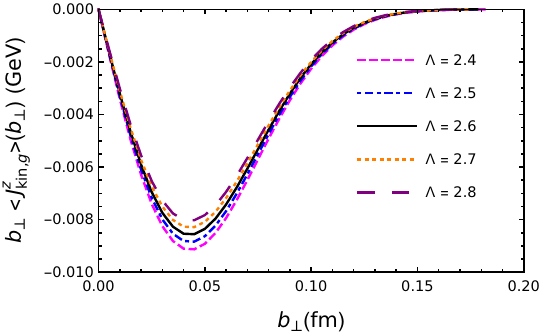}
    \end{minipage}

    \caption{Plot of the dependence of different components of AM distribution on transverse momentum UV cutoff. Top-left: Variation of $\langle L^z_{\text{can,g}} \rangle$ with $\Lambda$. Top-right: Variation of $\langle S^z_{\text{can,g}} \rangle$ with $\Lambda$. Bottom: Variation of $\langle J^z_{\text{can,g}} \rangle$ with $\Lambda$. Five different values of $\Lambda$ are considered for the analysis: $\Lambda = 2.4,2.5,2.6,2.7,2.8$ GeV.}
    \label{Cutoff_graph}
\end{figure}

In this section, we show the plots of the longitudinal component of the angular momentum distribution of gluons. The analysis in the previous section shows that the AM distribution for gluons is the same for Belinfante and Ji decomposition. So, we only plot the results for JM and Ji decomposition. As we stated before, we use a Gaussian wave packet state in transverse momentum space with fixed $p^+$.
For the analysis, we have chosen the parameters: the quark mass $m = 0.3$ GeV, the coupling constant $g = 1$, the color factor $C_f = 1$ and the ultraviolet cutoff $\Lambda = 2.63$ GeV. Here we show the variation of the angular momentum with respect to $|\vec{b}_\perp |$. Also, the y-axis is multiplied by a factor of $|\vec{b}_\perp |$ to correctly represent the data in radial coordinates. 

In Fig. \ref{gDist_graph}, the plot on the left panel is a graphical representation of $\langle J^z_{\text{can,g}} \rangle (b_\perp) = \langle L^z_{\text{can,g}} \rangle (b_\perp) + \langle S^z_{\text{can,g}} \rangle (b_\perp)$. The OAM has a positive contribution whereas the spin density has a negative contribution. 
In the right panel, we show that the kinetic (also valid for Belinfante) total AM distribution, $\langle J^z_{\text{kin,g}} \rangle (b_\perp)$ is not equal to the total AM density of the gluon. 
A correction term corresponding to the superpotential, $\langle M^z_{\text{g}} \rangle (b_\perp)$, which is ignored in a symmetric gluon part of the EMT like the Belinfante or kinetic decomposition, has to be added to 
$\langle J^z_{\text{kin,g}} \rangle (b_\perp)$ to get the same total AM distribution for both the decompositions. It is also evident from the comparison between these plots and those of the previous work that the gluon contribution to the AM density is significantly smaller than the quark contribution for the state that we have considered.

\begin{figure}[ht]
    \centering

    \begin{minipage}{0.45\textwidth}
        \centering
        \includegraphics[scale=0.9]{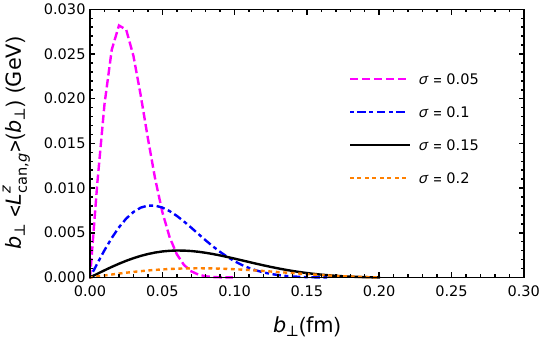}
    \end{minipage}
    \hfill
    \begin{minipage}{0.45\textwidth}
        \centering
        \includegraphics[scale=0.9]{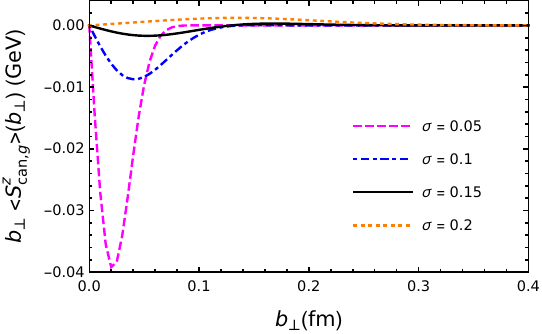}
    \end{minipage}

    \medskip

    \begin{minipage}{0.45\textwidth}
        \centering
        \includegraphics[scale=0.9]{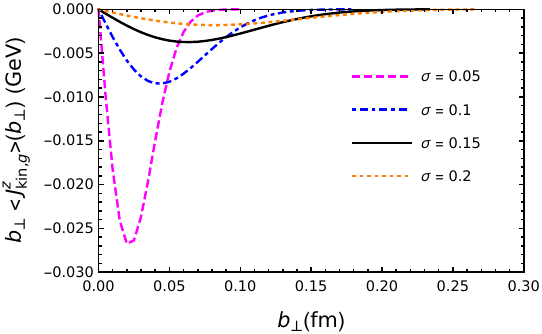}
    \end{minipage}
    \hfill
    \begin{minipage}{0.45\textwidth}
        \centering
        \includegraphics[scale=0.9]{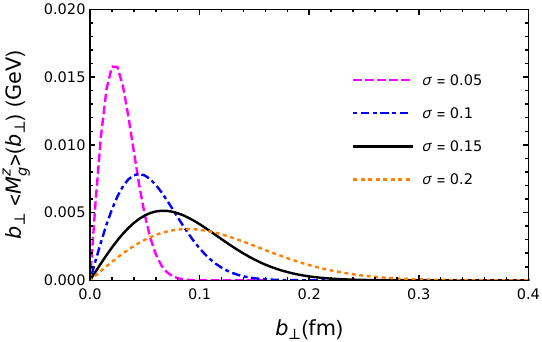}
    \end{minipage}

    \caption{Plots showing the dependence of different components of AM distribution on the width of the Gaussian wave-packet, $\sigma$. Four different values of $\sigma$ are considered for the analysis: $\sigma = 0.05,0.10,0.15,0.20$ GeV. Here, $m=0.3$ GeV, $\Lambda=2.63$ GeV, and $g=C_f=N_f=1$.}
    \label{gaussian_graph}
\end{figure}

As we discussed earlier, in systems with non-trivial interactions, such as in the dressed quark state, the quark and gluon contributions to the Ji angular momentum are dependent on the renormalization scale \cite{Ji:1995cu, Ji:2010zza, Altarelli:1977zs}.  In our framework, the scale dependence arises from imposing a cutoff on the integration over transverse momentum \cite{More:2021stk, More:2023pcy}.
In Fig. \ref{Cutoff_graph}, we show the cutoff dependence of different terms of the gluon AM density in our model. The correction terms used to compare the three definitions are cutoff-independent. In the previous work where we calculated the spatial densities of quark angular momentum, the expected equality $J_{\text{kin,q}}^{z}=J^{z}_{\text{Bel,q}}+M^{z}_{\text{q}}$ did not hold for different values of $\Lambda$. 
So we chose a suitable value for the cut-off to get the equality. However, in the case of gluon, the equality, $J_{\text{can,g}}^{z}=J^{z}_{\text{kin,g}}+M^{z}_{\text{g}}=J^{z}_{\text{Bel,g}}+M^{z}_{\text{g}}$, holds irrespective of the value of cutoff. Thus, the spatial density of the total angular momentum of gluon is the same for all the decompositions irrespective of the value of cutoff. However, the individual terms of the decompositions are still cutoff-dependent as expected. Fig. \ref{gaussian_graph} shows the variation of the results with the width of the wave packet state, $\sigma$. The width indicates the spread of the distribution.  We observe that in all cases, the peaks of the distributions shift away from the center, and the distributions become broader in $b_\perp$  space with an increase of the width of the Gaussian.

\section{Conclusions}
In this work, we have conducted a detailed investigation into the spatial distribution of angular momentum using various decomposition methods. We demonstrate that boundary terms, which vanish upon integration, play a crucial role when analyzing distributions in position space. Previous studies addressing spatial distributions of angular momentum components have predominantly relied on models of nucleon states without the inclusion of gluons. In contrast, our study employs a relativistic spin-1/2 state consisting of a quark dressed with a gluon, treated perturbatively within the one-loop approximation in QCD. This state incorporates gluonic degrees of freedom and enables the analytical calculation of quark-gluon light-front wave functions (LFWFs) using light-front Hamiltonian QCD. These LFWFs are frame-independent due to their dependence on internal coordinates.
Using a two-component formalism in the light-front gauge, we examine the distributions of longitudinal components of gluon spin, OAM, and total angular momentum, all derived from the overlaps of these two-particle LFWFs. 
Our study explicitly presents the gluonic contributions to the EMT, complementing previous work that focused on quark contributions to these observables. Furthermore, we illuminate the distinctions between the Ji and Jaffe-Manohar angular momentum decompositions coming from the superpotential term that becomes significant at the distribution level. We compute the missing superpotential term in the context of the dressed quark state. Additionally, in the light-front gauge, where the physical gauge field coincides with the total gauge potential, the decompositions by Wakamatsu and Chen et al. yield results consistent with Ji and JM decompositions. Lastly, we explicitly verify the longitudinal spin sum rule for all decomposition, we show that the contribution from the single particle sector of the Fock space expansion of the state plays an important role in the spin sum rule. We also explicitly show that by adding the quark and gluon contributions, the scale dependence in the total spin is canceled, as expected. 


\section{Acknowledgement}
 A. M. would like to thank SERB-POWER Fellowship (SPF/2021/000102) and SERB-MATRICS (MTR/2021/000103)  for financial support.
\appendix

 \section{Integrals used to calculate quark terms}\label{appA}	
	
	The following integrals are used to calculate the analytical expressions of the spatial distributions.
	
	\begin{align*}
		&\int d^2\kappa^{\perp} \frac{1}{D_1} = \pi \log\left[\frac{\Lambda^2+ m^2(1-x)^2}{m^2(1-x)^2}\right] , \tag{A1}\\
		&\int d^2\kappa^{\perp} \frac{1}{D_1 D_2} = \frac{\pi}{(1-x)^2} \frac1{q^2}\ \frac{f_2}{ f_1} , \tag{A2}\\
		&\int d^2\kappa^{\perp} \frac{\kappa^{(i)}}{D_1 D_2} = -\frac{\pi}{(1-x)}\frac{ q^{(i)}}{q^2 }\frac{f_2}{2 f_1}, \tag{A3}\\
		&\int d^2\kappa^{\perp} \frac{\kappa^{(1)}\kappa^{(2)}}{D_1 D_2}= \pi\frac{q^{(1)}q^{(2)}}{q^2}\bigg[-1+\left(1+\frac{2m^2}{q^2}\right)\frac{f_3}{2f_1}\bigg], \tag{A4}\\
		&\int d^2\kappa^{\perp} \frac{(\kappa^{(i)})^2}{D_1D_2} = \pi\bigg[-f_1 \ f_3+\frac{1}{2}+\frac{(q^{(i)})^2}{q^2}\left[\left(1+\frac{2m^2}{q^2}\right)\frac{f_3}{2f_1}-1\right]+\frac{1}{2}\log\bigg(\frac{\Lambda^2}{m^2(1-x)^2}\bigg),\bigg], \tag{A5}\\
		&\int d^2\kappa^{\perp} \frac{(\kappa^{(i)})^3}{D_1D_2} = \pi (1-x) \bigg[- \frac{3}{4} q^{(i)} \bigg[ 1+\log{\bigg(\frac{\Lambda^2}{m^2(1-x)^2}\bigg)} \bigg] + \frac{3}{2}\frac{(q^{(i)})^3}{q^2} + \bigg( \frac{6f_1}{4}q^{(i)} - \frac{(q^{(i)})^3}{q^2} \frac{4f_1^2-\frac{m^2}{q^2}}{2f_1} \bigg) f_3 \bigg], \tag{A6}\\
		&\int d^2\kappa^{\perp} \frac{(\kappa^{(1)})^2\kappa^{(2)}}{D_1D_2} = \pi (1-x) \bigg[- \frac{1}{4} q^{(2)} \bigg[ 1+\log{\bigg(\frac{\Lambda^2}{m^2(1-x)^2}\bigg)} \bigg] + \frac{3}{2}\frac{(q^{(1)})^2 q^{(2)}}{q^2} \bigg. \\
		& \hspace{9.95cm}+ \bigg( \frac{2f_1}{4}q^{(2)} - \frac{(q^{(1)})^2 q^{(2)}}{q^2} \frac{4f_1^2-\frac{m^2}{q^2}}{2f_1} \bigg) f_3 \bigg], \tag{A7}\\
		&\int d^2\kappa^{\perp} \frac{\kappa^{(1)}(\kappa^{(2)})^2}{D_1D_2} = \pi (1-x) \bigg[- \frac{1}{4} q^{(1)} \bigg[ 1+\log{\bigg(\frac{\Lambda^2}{m^2(1-x)^2}\bigg)} \bigg] + \frac{3}{2}\frac{q^{(1)} (q^{(2)})^2}{q^2} \\
		&\hspace{9.95cm}+ \bigg( \frac{2f_1}{4}q^{(1)} - \frac{q^{(1)}( q^{(2)})^2}{q^2} \frac{4f_1^2-\frac{m^2}{q^2}}{2f_1} \bigg) f_3 \bigg]. \tag{A8}
	\end{align*}
	where, $i= (1,2)$ and
	\begin{align*}
		D_1:= \bskasq +m^2(1-x)^2,\,\,\,\,\,\,\, D_2:=\left(\bska +(1-x)\bsq\right)^2+m^2(1-x)^2, \tag{A9}\\	f_1 :=\frac{1}{2}\sqrt{1+\frac{4 m^2}{q^2}}, \,\,\,\,\,\,\,\,\,f_2:=\log\left(1+\frac{q^2\left(1+2f_1\right)}{2 m^2}\right), \,\,\,\,\,\,\,
		f_3:= \log\left(\frac{1+2 f_1}{-1+2 f_1}\right). \tag{A10}
	\end{align*}

\section{Integrals used to calculate gluon terms}\label{appB}
	The following integrals are used to calculate the analytical expressions of the spatial distributions.
\begin{align}
\int_{-\infty}^{\infty} d^2\kappa^{\perp} \frac{1}{d_1}  =&~ \pi \ln \left[1+\frac{\Lambda^2}{m^2x^2}\right],\\
\int_{-\infty}^{\infty}d^2\kappa^{\perp}\, \frac{1}{d_1\ d_2}=&~\frac{2\pi}{\Delta^{\perp2}\left(1-x\right)^2}\frac{\tilde{f_2}}{\tilde{f_1}},\\
\int_{-\infty}^{\infty}d^2\kappa^{\perp}\, \frac{\kappa^{(i)}}{d_1\ d_2}=&~-\frac{\pi \Delta^{(i)}}{\Delta^{\perp2}\left(1-x\right)}\frac{\tilde{f_2}}{\tilde{f_1}},\\
\int_{-\infty}^{\infty}d^2\kappa^{\perp}\, \frac{\kappa^{(1)}\kappa^{(2)}}{d_1\ d_2}=&~\frac{\pi \Delta^{(1)}\Delta^{(2)}}{\Delta^{\perp2}}\left[-1+\left(\frac{1+\tilde{f_1}^2}{2\tilde{f_1}}\right)\tilde{f_2}\right],\\
\int_{-\infty}^{\infty}d^2\kappa^{\perp}\, \frac{(\kappa^{(i)})^2}{d_1\ d_2}=&~\pi \left[-\frac1{2}\tilde{f_1} \tilde{f_2}+\frac{1}{2}+\frac{ (\Delta^{(i)})^2}{\Delta^{\perp2}}\left(-1+\left(\frac{1+\tilde{f_1}^2}{2\tilde{f_1}}\right)\tilde{f_2}\right)\right]+\frac{\pi}{2} \ln\left(\frac{\Lambda^2}{m^2x^2}\right),\nnn  \\ 
\int_{-\infty}^{\infty}d^2\kappa^{\perp}\frac{\left(\kappa^{(i)}\right)^3}{d_1\ d_2}=& ~ \frac{3\pi}{4}\left(1-x\right)\Delta^{(i)}\left[-1+\tilde{f}_{2}-ln\left(\frac{\Lambda^2}{m^2x^2}\right)\right] + \pi \left(1-x\right)\frac{\left(\Delta^{(i)}\right)^{3}}{\Delta^{\perp2}}\left[\frac{3}{2}+\frac{1}{\tilde{f}_{1}}\left(1+\frac{3m^2x^2}{\left(1-x\right)^2\Delta^{\perp2}}\right)\tilde{f}_{2}\right],
 \\ \nonumber
\int_{-\infty}^{\infty}d^2\kappa^{\perp}\frac{\left(\kappa^{(1)}\right)^2\kappa^{(2)}}{d_1\ d_2}=& ~ \frac{\pi}{4}\left(1-x\right)\Delta^{(2)}\left[-1+\tilde{f}_{2}-ln\left(\frac{\Lambda^2}{m^2x^2}\right)\right]\\& + \pi \left(1-x\right)\frac{\left(\Delta^{(1)}\right)^{2}\Delta^{(2)}}{\Delta^{\perp2}}\left[\frac{3}{2}+\frac{1}{\tilde{f}_{1}}\left(1+\frac{3m^2x^2}{\left(1-x\right)^2\Delta^{\perp2}}\right)\tilde{f}_{2}\right],
\\ \nonumber
\int_{-\infty}^{\infty}d^2\kappa^{\perp}\frac{\kappa^{(1)}\left(\kappa^{(2)}\right)^{2}}{d_1\ d_2}=& ~ \frac{\pi}{4}\left(1-x\right)\Delta^{(1)}\left[-1+\tilde{f}_{2}-ln\left(\frac{\Lambda^2}{m^2x^2}\right)\right]\\& + \pi \left(1-x\right)\frac{\Delta^{(1)}\left(\Delta^{(2)}\right)^{2}}{\Delta^{\perp2}}\left[\frac{3}{2}+\frac{1}{\tilde{f}_{1}}\left(1+\frac{3m^2x^2}{\left(1-x\right)^2\Delta^{\perp2}}\right)\tilde{f}_{2}\right],
\end{align}
where, 
\begin{align}
d_1:=\left[\kappa^{\perp2}+m^2x^2\right],\,\,\,\,\,\,
d_2:=\left[\left(\bska+\left(1-x\right)\bsD \right)^2+m^2x^2\right],\\
\tilde{f_1}:= \sqrt{1+\frac{4m^2x^2}{q^{\perp2}\left(1-x\right)^2}},\,\,\,\,\,\,\,\,\,
\tilde{f_2}:= \ln\left(\frac{1+\tilde{f_1}}{-1+\tilde{f_1}}\right).    
\end{align}

\section{Spatial distribution of angular momentum components of kinetic (Ji) decomposition}\label{appC}
	\subsection{Quark OAM distribution}
	 The distribution of kinetic quark OAM has already been derived in the previous work. We just state the final result here for completion 
	\begin{align}
		\langle L^z_{\text{kin,q}} \rangle (\bsb) &= i \int \frac{d^2\boldsymbol{\Delta}^{\perp}}{(2\pi)^2}e^{-i\bsb \cdot \bsD}\left[ \frac{\partial \langle T^{+1}_{\text{kin,q}}\rangle_{\text{LF}} }{\partial \Delta_{\perp}^{(2)}}-\frac{\partial \langle T^{+2}_{\text{kin,q}}\rangle_{\text{LF}} }{\partial \Delta_{\perp}^{(1)}}\right]_{\text{DY}}. \nonumber \\
		&=\frac{g^2 C_F}{72 \pi^2} \int \frac{d^2\boldsymbol{\Delta}^{\perp}}{(2\pi)^2}e^{-i\bsb \cdot \bsD} \left[-7+\frac{6}{\omega}\left(1+\frac{2m^2}{\Delta^2}\right)\log\left(\frac{1+\omega}{-1+\omega}\right)-6~\log\left(\frac{\Lambda^2}{m^2}\right)\right], \label{Lzkin}
	\end{align}
	where $\omega = \sqrt{1+\frac{4m^2}{\Delta^2}}$.

	\subsection{Quark spin distribution}
	Consider the quark spin operator from Eq. \eqref{J_kin} 
	\begin{align}
 S^{+jk}_{\text{kin,q}} &= \frac{1}{2} \epsilon^{+jk-}\sum_{\lambda,\lambda^{\prime}}\int \frac{dk^{\prime+}d^2\boldsymbol{k}^{\prime \perp}dk^+d^2\boldsymbol{k}^{\perp}}{\left(16 \pi\right)^2 \sqrt{k^{\prime +}k^+}}b_{\lambda^{\prime}}^{\dagger}(k^{\prime})b_{\lambda}(k) ~ \left(\chi_{\lambda^{\prime}}^{\dagger}\sigma^{(3)}\chi_{\lambda}\right).
	\end{align}
\textbf{\underline{Non-diagonal matrix element of $S_{\text{kin,q}}^{+jk}(0)$:}}
	\begin{align}
		\langle 1|S_{\text{kin,q}}^{+jk}(0)|2 \rangle = \langle 2|S_{\text{kin,q}}^{+jk}(0)|1 \rangle = 0,
	\end{align}	
	\textbf{\underline{Diagonal matrix element of $S_{\text{kin,q}}^{+jk}(0)$:}}\\
 
 The single-particle contribution is given as  
 \begin{align}
     \frac{\langle 1,\uparrow |S_{\text{kin,q}}^{+jk}(0)|1,\uparrow \rangle}{2p^+} &=  \frac{\epsilon{+jk-}}{4} |\psi_1|^2  = \frac{\epsilon^{+jk-}}{4} \left[ 1-\frac{g^2}{8 \pi^2}C_f \int dx \left\{\frac{1+x^2}{1-x} \log{\left(\frac{\Lambda^2}{m^2 (1-x)^2}\right)} -\frac{2x}{1-x}\right\}\right]. \label{SME1}
 \end{align}
 The two-particle contribution is
	\begin{align}
		&\frac{\langle 2,\uparrow |S_{\text{kin,q}}^{+jk}(0)|2,\uparrow \rangle}{2p^+}  = \frac{g^2C_f}{4}\epsilon^{+jk-}\int \frac{dxd^2\bska}{8\pi^3} \frac{1}{(1-x)D_1D_2}\times \nonumber \\
		& \hspace{0.4cm} \left[ {\kappa^{\perp2}} (1 + x^2) + \bska \cdot \bsD (1 - x)(1 + x^2) + i(1-x)(1 - x^2) (\kappa^{(1)}\Delta^{(2)}-\kappa^{(2)}\Delta^{(1)}) - m^2(1 - x)^4\right]. \label{SME2}
	\end{align}
	After performing the $\kappa$ integration in the above equation using Appen. \ref{appA} and using Eq. \eqref{Sdist}, we get:
	\begin{align}
		&\langle S^z_{\text{kin,q}} \rangle (\bsb) = -
  \frac{1}{2}\epsilon^{3jk}\int \frac{d^2\bsD}{(2\pi)^2}e^{-i \bsD \cdot \bsb} \langle S^{+jk} \rangle_{\text{LF}} \bigg|_{\text{DY}}\nonumber \\ &=- \frac{\epsilon^{3jk} \epsilon^{+jk-}}{2} \int \frac{d^2\bsD}{(2\pi)^2}e^{-i \bsD \cdot \bsb}  \int dx \left[ \frac{1}{4}-\frac{g^2 C_f}{32 \pi^2} \left\{\frac{1+x^2}{1-x} \ln{\left(\frac{\Lambda^2}{m^2 (1-x)^2}\right)} -\frac{2x}{1-x}\right\}\right. \nonumber \\
		&\left.-\frac{g^2 C_f}{32\pi^2(1-x)}\left\{ \omega (1+x^2) \log{\left( \frac{1+\omega}{-1+\omega} \right)} + \left(\frac{1-\omega^2}{\omega}\right) x \log{\left( \frac{1+\omega}{-1+\omega} \right)} - (1+x^2) \log{\left( \frac{\Lambda^2}{m^2(1-x)^2} \right)} \right\} \right] \nonumber 
	 \\
 &= \int \frac{d^2\bsD}{(2\pi)^2}e^{-i \bsD \cdot \bsb}  \int dx \left[ \frac{1}{2}+\frac{g^2 C_f}{16 \pi^2(1-x)} \left\{2x- \left[\omega (1+x^2) + \left(\frac{1-\omega^2}{\omega}\right) x \right] \log{\left( \frac{1+\omega}{-1+\omega} \right)} \right\} \right], \hspace{1cm}\label{Szkin}
	\end{align}
\subsection{Gluon total AM distribution}
 The operator structure of gluon EMT in the kinetic decomposition is given by Eq.\eqref{T_kin}
 \begin{align}
     \nonumber T^{+k}_{\text{kin,g}}=& -2 \text{Tr}\left[G^{+\lambda}G^{k}_{\lambda}\right] \\ \nonumber =& 2\text{Tr}\left[\frac{1}{2}G^{+-}G^{+k}+G^{+\perp}G^{k\perp}\right]
   \\  =& \frac{1}{2}G_{a}^{+-}G_{a}^{+k}+G_{a}^{+\perp}G_{a}^{k\perp}\\
   \nonumber =& \underbrace{\left(\partial^{+}A_{a}^{k}\right)\left(\partial^{\perp}A_{a}^{\perp}\right)+\left(\partial^{+}A_{a}^{\perp}\right)\left(\partial^{k}A_{a}^{\perp}\right)-\left(\partial^{+}A_{a}^{\perp}\right)\left(\partial^{\perp}A_{a}^{k}\right)}_{\text{D}}+\underbrace{2g\left(\partial^{+}A_{a}^{k}\right)\left(\frac{1}{\partial^{+}}\right)\left(\xi^{\dagger}t^{a}\xi\right) }_{\text{ND}}
 \end{align}
The notations D and ND in the above expression denote the diagonal and non-diagonal overlap contributions to the matrix elements, respectively. We ignore the higher-order terms containing three gluon fields for our analysis.\\

\textbf{\underline{Diagonal matrix element of $T^{+k}_{\text{kin,g}}(0)$:}}\\

Substituting the expressions of gluon fields in the diagonal terms we get 
\begin{align}
&\left(\partial^{+}A_{a}^{k}\right)\left(\partial^{\perp}A_{a}^{\perp}\right)+\left(\partial^{+}A_{a}^{\perp}\right)\left(\partial^{k}A_{a}^{\perp}\right)-\left(\partial^{+}A_{a}^{\perp}\right)\left(\partial^{\perp}A_{a}^{k}\right) \nonumber 
\\ &= \sum_{\lambda, \lambda^{\prime}}\int \frac{dk^{+}d^2\bsk dk^{+}d^2\bskpr}{\left(16\pi^3\right)^2k^{\prime +}}\bigg[\left\{-\epsilon^{k}_{\lambda}\left(k^{\prime \perp}\cdot \epsilon^{\perp}_{\lambda^{\prime}}\right)-k^{\prime k}\left(\epsilon^{\perp}_{\lambda}\cdot \epsilon^{\perp}_{\lambda^{\prime}}\right)+\epsilon^{k}_{\lambda^{\prime}}\left(k^{\prime \perp}\cdot\epsilon^{\perp}_{\lambda}\right)\right\}a_{\lambda}(k)a_{\lambda^{\prime}}(k^{\prime})e^{-i\left(k+k^{\prime}\right)\cdot y} \nonumber \\ & 
+ \left\{\epsilon^{k}_{\lambda}\left(k^{\prime \perp}\cdot \epsilon^{\perp*}_{\lambda^{\prime}}\right)+k^{\prime k}\left(\epsilon^{\perp}_{\lambda}\cdot \epsilon^{\perp*}_{\lambda^{\prime}}\right)-\epsilon^{k*}_{\lambda^{\prime}}\left(k^{\prime \perp}\cdot\epsilon^{\perp}_{\lambda}\right)\right\}a_{\lambda}(k)a^{\dagger}_{\lambda^{\prime}}(k^{\prime})e^{-i\left(k-k^{\prime}\right)\cdot y} \nonumber \\ & 
+ \left\{\epsilon^{k*}_{\lambda}\left(k^{\prime \perp}\cdot \epsilon^{\perp}_{\lambda^{\prime}}\right)+k^{\prime k}\left(\epsilon^{\perp*}_{\lambda}\cdot \epsilon^{\perp}_{\lambda^{\prime}}\right)-\epsilon^{k}_{\lambda^{\prime}}\left(k^{\prime \perp}\cdot\epsilon^{\perp*}_{\lambda}\right)\right\}a^{\dagger}_{\lambda}(k)a_{\lambda^{\prime}}(k^{\prime})e^{i\left(k-k^{\prime}\right)\cdot y} \nonumber \\ & 
+ \left\{-\epsilon^{k*}_{\lambda}\left(k^{\prime \perp}\cdot \epsilon^{\perp*}_{\lambda^{\prime}}\right)-k^{\prime k}\left(\epsilon^{\perp*}_{\lambda}\cdot \epsilon^{\perp*}_{\lambda^{\prime}}\right)+\epsilon^{k*}_{\lambda^{\prime}}\left(k^{\prime \perp}\cdot\epsilon^{\perp*}_{\lambda}\right)\right\}a^{\dagger}_{\lambda}(k)a^{\dagger}_{\lambda^{\prime}}(k^{\prime})e^{i\left(k^{\prime}+k\right)\cdot y}\bigg] \hspace{4.6cm}
\end{align}
The diagonal contribution coming from this operator is given by
\begin{align}
   &\frac{\langle 2,\uparrow|T^{+k}_{\text{kin,g}}|2,\uparrow \rangle}{2p^+} \nonumber \\ =& \frac{1}{2}\sum_{\lambda_{1},\lambda_{2},\lambda_{2}^{\prime}}\int dx d^2\boldsymbol{\kappa}^{\perp} \phi^{\sigma^{\prime}*}_{\lambda_{1},\lambda_{2}^{\prime}}\left(x, \bs{\kappa}^{\prime\perp}\right)\bigg[\epsilon^{k}_{\lambda_{2}}\left(\bs{\kappa}^{\prime \perp}+x\frac{\bs{\Delta}^{\perp}}{2}\right)\cdot \bs{\epsilon}^{\perp*}_{\lambda_{2}^{\prime}}+\epsilon^{k*}_{\lambda_{2}^{\prime}}\left(\bs{\kappa}^{ \perp}-x\frac{\bs{\Delta}^{\perp}}{2}\right)\cdot \bs{\epsilon}^{\perp}_{\lambda_{2}} \nonumber\\\nonumber &+\left({\kappa}^{\prime k}+x\frac{\Delta^{k}}{2}\right)\left(\bs{\epsilon}^{\perp}_{\lambda_{2}} \cdot \bs{\epsilon}^{\perp*}_{\lambda_{2}^{\prime}}\right)+\left({\kappa}^{ k}-x\frac{\Delta^{k}}{2}\right)\left(\bs{\epsilon}^{\perp*}_{\lambda_{2}^{\prime}} \cdot \bs{\epsilon}^{\perp}_{\lambda_{2}}\right) -\epsilon^{k*}_{\lambda^{\prime}_{2}}\left(\bs{\kappa}^{\prime \perp}+x\frac{\bs{\Delta}^{\perp}}{2}\right)\cdot \bs{\epsilon}^{\perp}_{\lambda_{2}}\\&-\epsilon^{k}_{\lambda_{2}}\left(\bs{\kappa}^{\perp}-x\frac{\bs{\Delta}^{\perp}}{2}\right)\cdot \bs{\epsilon}^{\perp*}_{\lambda_{2}^{\prime}}\bigg]\phi^{\sigma}_{\lambda_{1},\lambda_{2}}\left(x,\bs{\kappa}^{\perp}\right). \hspace{9.5cm}
   \end{align}
By substituting the two-particle LFWF in the above expression, evaluating the spinor products, and performing the kappa integrals gives the following expression for the components of the matrix element of gluon EMT
   \begin{align}
 \nonumber  & \frac{\langle 2, \uparrow|T^{+1}_{\text{kin, g}}(0)|2, \uparrow\rangle}{2p^{+}} \\\nonumber =& g^{2}C_{f}\int \frac{dxd^{2}\bska}{16 \pi^{3}}\frac{1}{x d_{1}d_{2}}\bigg[\kappa^{(1)3}\left(4-2x(2-x)\right)+\kappa^{(1)}\kappa^{(2)2}\left(4-2x(2-x)\right)+m^{2}x^{4}\left(\Delta^{(1)}-x\Delta^{(1)}-i\Delta^{(2)}\right) \\ \nonumber & \kappa^{(2)2}\left((1-x)(2-(2-x)x)\Delta^{(1)}-i(2-x)x\Delta^{(2)}\right)+\kappa^{(1)}\kappa^{(2)}\left(2(1-x)(2\Delta^{(2)}-i(2-x)x(\Delta^{(1)}-i\Delta^{(2)}))\right)  \bigg]     \\ =& ig^{2}C_{f}\int \frac{dx}{16 \pi^{2}}\Delta^{(2)}\bigg[-\frac{x\left(4m^{2}x-\left(2-x\right)\Delta^{2}\right)\log\left(\frac{1+\omega'}{-1+\omega'}\right)}{\Delta^{2}\omega'} + \left(2-x\right)\left(1-x-x\log\left(\frac{\Lambda^{2}}{m^{2}x^{2}}\right)\right)  \bigg] \label{T+1king}
\end{align}

\begin{align}
 \nonumber  & \frac{\langle 2, \uparrow|T^{+2}_{\text{kin, g}}(0)|2, \uparrow\rangle}{2p^{+}} \\ =& ig^{2}C_{f}\int \frac{dx}{16 \pi^{2}}\Delta^{(1)}\bigg[\frac{x\left(4m^{2}x-\left(2-x\right)\Delta^{2}\right)\log\left(\frac{1+\omega'}{-1+\omega'}\right)}{\Delta^{2}\omega'} - \left(2-x\right)\left(1-x-x\log\left(\frac{\Lambda^{2}}{m^{2}x^{2}}\right)\right)  \bigg] 
 \hspace{1cm} \label{T+2king}
\end{align}
where $\omega'=\sqrt{1+\frac{4m^{2}x^{2}}{\left(1-x\right)^{2}\Delta^{2}}}$ and the kappa integrals are given in the Appn. \ref{appB}\\

\textbf{\underline{Non-diagonal matrix element of $T^{\mu\nu}_{\text{kin, g}}(0)$}}\\

The operator structure of the non-diagonal component is 
\begin{align}
   & \nonumber 2g\left(\partial^{+}A^{k}_{a}\right)\left(\frac{1}{\partial^{+}}\right)\left(\xi^{\dagger}t^{a}\xi\right) \\ =& 2g \sum_{\lambda^{\prime},\lambda, \lambda_{3}}\int \frac{dk^{+}d^2\bs{k}^{\perp}dk^{\prime+}d^2\bs{k}^{\prime\perp}dk_{3}^{+}d^2\bs{k}_{3}^{\perp}}{\left(16\pi^{3}\right)^{3}\sqrt{k^{\prime +}k^{+}}}\left[\epsilon^{k}_{\lambda_{3}}a_{\lambda_{3}}(k_{3})b_{\lambda^{\prime}}^{\dagger}(k^{\prime})b_{\lambda}(k)- \epsilon^{k*}_{\lambda_{3}}a^{\dagger}_{\lambda_{3}}(k_{3})b_{\lambda^{\prime}}^{\dagger}(k^{\prime})b_{\lambda}(k)\right]\frac{\chi^{\dagger}_{\lambda^{\prime}}t^a\chi_{\lambda}}{k^{+}-k^{\prime +}}.
\end{align}
The contribution to the matrix element coming from this operator is given by 
\begin{align}
   \nonumber & \frac{1}{2p^{+}}\left[\langle 1, \sigma^{\prime}|2g\left(\partial^{+}A^{k}_{a}\right)\left(\frac{1}{\partial^{+}}\right)\left(\xi^{\dagger}t^{a}\xi\right)|2, \sigma \rangle + \langle 2, \sigma^{\prime}|2g\left(\partial^{+}A^{k}_{a}\right)\left(\frac{1}{\partial^{+}}\right)\left(\xi^{\dagger}t^{a}\xi\right)|1, \sigma \rangle\right] \\=& \frac{g}{\sqrt{16\pi^{3}}}\sum_{\lambda_{1},\lambda_{2}}\int dx d^2\kappa^{\perp}\left[\left(-\frac{\epsilon^{k}_{\lambda_{2}}}{\sqrt{x}}\right)\left(\chi^{\dagger}_{\sigma^{\prime}}t^{a}\chi_{\lambda_{1}}\right)\phi^{\sigma}_{\lambda_{1},\lambda_{2}}\left(x, \bska\right)+\phi^{\sigma^{\prime*}}_{\lambda_{1},\lambda_{2}}\left(x, \bska\right)\left(\chi^{\dagger}_{\lambda_{1}}t^{a}\chi_{\sigma}\right)\left(-\frac{\epsilon^{k*}_{\lambda_{2}}}{\sqrt{x}}\right)\right],
\end{align}
where $x$ and $\bs{\kappa}^{\perp}$ are the longitudinal and transverse momentum fraction of the gluon inside the dressed quark state. This spinor product gives an integrand that is odd in $\kappa$. Thus, after $\kappa$ integration, this term vanishes we have no contribution from the off-diagonal terms. So, we only have to evaluate the contribution of diagonal terms
We evaluate the contribution to $\langle T^{+k}_{\text{kin,g}} \rangle_{\text{LF}}$ from the diagonal term i.e. Eq.\eqref{T+1king} and \eqref{T+2king}. 
	\begin{align}
		\langle J^z_{\text{kin,g}} \rangle (\bsb)= i \int \frac{d^2\boldsymbol{\Delta}^{\perp}}{(2\pi)^2}e^{-i\bsb \cdot \bsD}\left[ \frac{\partial \langle T^{+1}_{\text{kin,g}}\rangle_{\text{LF}} }{\partial \Delta_{\perp}^{(2)}}-\frac{\partial \langle T^{+2}_{\text{kin,g}}\rangle_{\text{LF}} }{\partial \Delta_{\perp}^{(1)}}\right]_{\text{DY}}. \label{Jzg_full}
	\end{align}
From Eq.\eqref{Jzg_full} we get: 
	\begin{align}
		\langle J^z_{\text{kin,g}} \rangle (\bsb)&= g^2 C_f \int \frac{d^2\boldsymbol{\Delta}^{\perp}}{(2\pi)^2}e^{-i\bsb \cdot \bsD} \int \frac{dx}{16 \pi^2} \frac{1}{\left(1-x\right)^2{\Delta}^4 \omega'^3} ~\times \nonumber \\
&\left[x \left(8 m^4 x^3 + 6 m^2 (x-2) x^2 \Delta^2 + (x-2) (x-1)^2 \Delta^4\right) \log\left(\frac{1 + \omega'}{-1 + \omega'}\right)\right. \nonumber \\
&\left.+ \Delta^2 \omega' \left((x-1) \left(4 m^2 x^2 + (x-2) (x-1 ) \Delta^2\right) - (x-2) x \left(4 m^2 x^2 + (x-1)^2 \Delta^2\right) \log\left(\frac{\Lambda^2}{m^2 x^2}\right)\right) \right] \label{Jzkin}
	\end{align}

\subsection{Spatial distribution of superpotential for quark}
The distribution of the correction term corresponding to the superpotential in Eq. \eqref{kin&Bel} can be evaluated using Eq. \eqref{Mzgeneral} with $\kappa=1/2$ and  $S^{l+k}_q$
\begin{align}
   \langle M^{z}_q\rangle (\bsb)= \frac{1}{2}\epsilon^{3jk} \int \frac{d^2 \bsD}{(2\pi)^{2}}~e^{i \boldsymbol{\Delta}_\perp \cdot \boldsymbol{b}_\perp}  ~\Delta^{l}\frac{\partial }{\partial \Delta^{j}} \langle S^{l+k}_{\text{kin,q}}  \rangle ,  \label{Mzq_dist}
   \end{align}
\textbf{\underline{Diagonal matrix element of $S_{\text{kin,q}}^{l+k}(0)$:}}\\

Using Eq. \eqref{SME1} and \eqref{SME2} with $\kappa$ integration carried out, we get
\begin{align}
\langle S_{\text{kin,q}}^{l+k}& \rangle = \frac{\langle 1|S_{\text{kin,q}}^{l+k}(0)|1 \rangle}{2p^+} + \frac{\langle 2|S_{\text{kin,q}}^{l+k}(0)|2 \rangle}{2p^+} = \epsilon^{l+k-} \left[ \frac{1}{4}-\frac{g^2}{32 \pi^2}C_f \int dx \left\{\frac{1+x^2}{1-x} \ln{\left(\frac{\Lambda^2}{m^2 (1-x)^2}\right)} -\frac{2x}{1-x}\right\}\right] \nonumber \\
&-\epsilon^{l+k-}\left[\frac{g^2 C_f}{32\pi^2}\int \frac{dx}{1-x}  \left\{ \left[\omega (1+x^2) + \left(\frac{1-\omega^2}{\omega}\right) x \right] \log{\left( \frac{1+\omega}{-1+\omega} \right)} - (1+x^2) \log{\left( \frac{\Lambda^2}{m^2(1-x)^2} \right)} \right\}\right].
\end{align}
\textbf{\underline{Non-diagonal matrix element of $S_{\text{kin,q}}^{l+k}(0)$:}}\\
	\begin{align}
		\langle 1|S_{\text{kin,q}}^{l+k}(0)|2 \rangle = \langle 2|S_{\text{kin,q}}^{l+k}(0)|1 \rangle = 0.
	\end{align}	
Substituting these expressions in Eq. \eqref{Mzq_dist}, we get

\begin{align}
\langle M^{z}_q\rangle (\bsb)
=& \frac{g^2C_{f}}{16 \pi^2}\int \frac{d^2\bsD}{(2\pi)^2}e^{-i\bsD \cdot \bsb}\int \frac{dx}{\left(1-x\right)\omega^3 \Delta^{4}}  \left[ \left(\omega \Delta^{2}-2m^2\right)\left(\left(4m^2+\Delta^{2}\right)\left(1+x^2\right)-4m^2x\right)+4m^2x\Delta^2\right] 
\end{align}

\section{Spatial distribution of angular momentum components of Belinfante (Bel) decomposition} \label{appD}
	\subsection{Quark total AM distribution}
Consider the quark part of Belinfante EMT in Eq. \eqref{TBel}
	\begin{align}
		T^{\mu\nu}_{\text{Bel,q}}(x)=& \frac{1}{4}\overline{\psi}(x)\left[\gamma^{\mu}i\overleftrightarrow{D}^{\nu}+\gamma^{\nu}i\overleftrightarrow{D}^{\mu}\right]\psi(x) \nonumber \\
	\therefore T^{+k}_{\text{Bel,q}} =& \underbrace{\frac{1}{2}T^{+k}_{\text{kin,q}}(x)}_{\text{1st term}}+\underbrace{\frac{1}{4}\overline{\psi}(x)\left[\gamma^{k}i \overleftrightarrow{\partial}^{+}\right]\psi(x)}_{\text{2nd term}},
	\end{align}
	the first term in this expression has already been calculated. 
	\begin{align}
		&\nonumber \text{2nd term has the following operator structure }\\\nonumber&= \frac{1}{4}\sum_{\lambda,\lambda^{\prime}}\int \frac{dk^{\prime+}d^2\boldsymbol{k}^{\perp}dk^{+}d^2\boldsymbol{k}^{\perp}}{\left(16\pi^3\right)^2\sqrt{k^{\prime+}k^+}}e^{ik^{\prime}\cdot y}\chi_{\lambda^{\prime}}^{\dagger}\left[\sigma^{k}\left(k^{+}+k^{\prime+}\right)\left(\frac{1}{k^{+}}\right)\left(\sigma^{i}k^{i}+im\right)+\left(\sigma^{i}k^{\prime i}-im\right)\left(\frac{1}{k^{\prime+}}\right)\sigma^{k}\left(k^{+}+k^{\prime +}\right)\right]\chi_{\lambda} \\\nonumber & \times e^{-i k \cdot y} b_{\lambda^{\prime}}^{\dagger}(k^{\prime})b_{\lambda}(k)+\frac{g}{4}\sum_{\lambda,\lambda^{\prime},\lambda_{3}}\int \frac{dk^{\prime+}d^2\boldsymbol{k}^{\perp}dk^{+}d^2\boldsymbol{k}^{\perp}dk_{3}^{+}d^2\boldsymbol{k_3}^{\perp}}{\left(16\pi^3\right)^3k_{3}^{+}\sqrt{k^{\prime+}k^+}}e^{ik^{\prime}\cdot y}\chi_{\lambda^{\prime}}^{\dagger}\bigg[\sigma^{k}\bigg\{\frac{k^{+}+k_3^{+}+k^{\prime +}}{k^{+}+k_3^{+}}\left(\sigma^{i}\epsilon_{\lambda_3}^{i}\right)a_{\lambda_3}(k_3)e^{-ik_3 \cdot y}\\\nonumber&+\frac{k^+-k_3^++k^{\prime +}}{k^{+}-k_3^{+}}\left(\sigma^{i}\epsilon_{\lambda_3}^{i*}\right)a^{\dagger}_{\lambda_3}(k_3)e^{ik_3 \cdot y} \bigg\} + \bigg\{\left(\sigma^{i}\epsilon_{\lambda_3}^{i}\right)\frac{k^{+}+k^{\prime +}-k_3^{+}}{k^{\prime +}-k_3^{+}}a_{\lambda_3}(k_3)e^{-ik_3 \cdot y}\\ &+\left(\sigma^{i}\epsilon_{\lambda_3}^{i*}\right)\frac{k^{+}+k^{\prime +}+k_3^+}{k^{\prime +}+k_3^{+}}a^{\dagger}_{\lambda_3}(k_3)e^{ik_3 \cdot y}\bigg\}\sigma^{k}\bigg]\chi_{\lambda}e^{-i k \cdot y}b_{\lambda^{\prime}}^{\dagger}(k^{\prime})b_{\lambda}(k).
	\end{align}
	
	\textbf{\underline{Diagonal matrix element of $T^{\mu\nu}_{\text{Bel,q}}(0)$:}}\\
 
	We know from the calculation of $T^{+k}_{\text{kin,q}}$ that it does not have any single-particle contribution; so the only single-particle contribution comes from the second term and it is a diagonal overlap :
 \begin{align}
     &\frac{\langle 1,\uparrow|T^{+k}_{\text{Bel,q}}|1,\uparrow \rangle}{2p^+} = \frac{1}{8} |\psi_1|^2 \chi_{\sigma'}^\dagger \left\{ \Delta^i \left( \sigma^i \sigma^k - \sigma^k \sigma^i  \right) \right\} \chi_\sigma \hspace{9cm} \nonumber \\
     &\frac{\langle 1,\uparrow|T^{+1}_{\text{Bel,q}}|1,\uparrow \rangle}{2p^+} = -\frac{i}{4} \left[|\psi_1|^2 \Delta^{(2)} \chi_{\sigma'}^\dagger \sigma^{(3)} \chi_\sigma \right] = -\frac{i}{4} |\psi_1|^2 \Delta^{(2)}  \label{T+1_Bel1}\\
     &\frac{\langle 1,\uparrow|T^{+2}_{\text{Bel,q}}|1,\uparrow \rangle}{2p^+} = \frac{i}{4} \left[|\psi_1|^2 \Delta^{(1)} \chi_{\sigma'}^\dagger \sigma^{(3)} \chi_\sigma \right] = \frac{i}{4} |\psi_1|^2 \Delta^{(1)} \label{T+2_Bel1} 
 \end{align}
	\begin{align}
		&\nonumber \frac{\langle 2,\uparrow|T^{+k}_{\text{Bel,q}}|2,\uparrow \rangle}{2p^+} \\\nonumber  &= \frac{1}{2}\sum_{\lambda_2,\lambda_1,\lambda_1^{\prime}}\int dx d^2\bska~ \phi^{*\sigma^{\prime}}_{\lambda_1^{\prime},\lambda_2}\left(x,\bskapr\right)\chi^{\dagger}_{\lambda_1^{\prime}}\bigg[\left(\boldsymbol{\kappa}^k+\left(1-x\right)\frac{\boldsymbol{\Delta}^k}{2}\right)+\left(\boldsymbol{\sigma}^{i}\left(\boldsymbol{\kappa}^{\prime i}+x\frac{\boldsymbol{\Delta}^{i}}{2}\right)-im\right)\frac{\boldsymbol{\sigma}^{k}}{2}\\&+\frac{\boldsymbol{\sigma}^{k}}{2}\left(\boldsymbol{\sigma}^{i}\left(\boldsymbol{\kappa}^{i}-x\frac{\boldsymbol{\Delta}^{i}}{2}\right)+im\right) \bigg]\chi_{\lambda_1}\phi^{\sigma}_{\lambda_1,\lambda_2}\left(x,\bska\right),  \hspace{8.9cm}
	\end{align}
	where $x$ is the quark momentum fraction and $\bskapr=\bska+(1-x)\bsD$.
	\begin{align}
		&\nonumber \frac{\langle 2,\uparrow|T^{+1}_{\text{Bel,q}}|2,\uparrow \rangle}{2p^+} \\\nonumber&= g^2C_F \int \frac{dxd^2\bska}{32\pi^3}\frac{1}{\left(1-x\right)}\frac{m^2\left(1-x\right)^4\left(4\kappa^{(1)}+2\left(1-x\right)\Delta^{(1)}+i\Delta^{(2)}\right)}{D_1D_2} ~\times\\& \frac{\left(4\kappa^{(1)}+2\left(1-x\right)\Delta^{(1)}-i\Delta^{(2)}\right)\left[\left(1+x^2\right)\left(\kappa^{\perp2}+\left(1-x\right)\bska \cdot \bsD\right)+i\left(1-x\right)\left(1-x^2\right)\left(\kappa^{(1)}\Delta^{(2)}-\kappa^{(2)}\Delta^{(1)}\right)\right]}{D_1D_2}
		\label{T+1_Bel2}. \\
		&\nonumber \frac{\langle 2,\uparrow|T^{+2}_{\text{Bel,q}}|2,\uparrow \rangle}{2p^+} \\\nonumber&= g^2C_F \int \frac{dxd^2\bska}{32\pi^3}\frac{1}{\left(1-x\right)}\frac{m^2\left(1-x\right)^4\left(4\kappa^{(2)}+2\left(1-x\right)\Delta^{(2)}-i\Delta^{(1)}\right)}{D_1D_2} ~\times\\& \frac{\left(4\kappa^{(2)}+2\left(1-x\right)\Delta^{(2)}+i\Delta^{(1)}\right)\left[\left(1+x^2\right)\left(\kappa^{\perp2}+\left(1-x\right)\bska \cdot \bsD\right)+i\left(1-x\right)\left(1-x^2\right)\left(\kappa^{(1)}\Delta^{(2)}-\kappa^{(2)}\Delta^{(1)}\right)\right]}{D_1D_2}.
		\label{T+2_Bel2}
	\end{align}
	\textbf{\underline{Non-diagonal matrix element of $T^{\mu\nu}_{\text{Bel,q}}(0)$:}}
	\begin{align}
		&\nonumber \frac{1}{2p^+}\left[\langle 1,\uparrow|T^{+k}_{\text{Bel,q}}|2,\uparrow \rangle + \langle 2,\uparrow|T^{+k}_{\text{Bel,q}}|1,\uparrow \rangle\right] \\ \nonumber =&\frac{g}{4\sqrt{16\pi^3}}\int dx d^2\bska~ \frac{1}{\sqrt{\left(1-x\right)}}\bigg[ \chi^{\dagger}_{\sigma^{\prime}}\left(\boldsymbol{\sigma}^{k}\left(\boldsymbol{\sigma}^{i}\boldsymbol{\epsilon}^{i}_{\lambda_2}\right)+\left(\boldsymbol{\sigma}^{i}\boldsymbol{\epsilon}^{i}_{\lambda_2}\right)\boldsymbol{\sigma}^{k} + 2 \boldsymbol{\epsilon}^{k}_{\lambda_2} \right)\chi_{\lambda_1}\phi_{\lambda_1,\lambda_2}^{\sigma}\left(x,\bska\right)\\ \nonumber&+\phi_{\lambda_1,\lambda_2}^{*\sigma^{\prime}}\left(x,\bska\right)\chi^{\dagger}_{\lambda_1}\left(\left(\boldsymbol{\sigma}^{i}\boldsymbol{\epsilon}^{i*}_{\lambda_2}\right)\boldsymbol{\sigma}^{k}+\boldsymbol{\sigma}^{k}\left(\boldsymbol{\sigma}^{i}\boldsymbol{\epsilon}^{i*}_{\lambda_2}\right) + 2 \boldsymbol{\epsilon}^{k*}_{\lambda_2} \right)\chi_{\sigma}\bigg].
	\end{align}
	This spinor product gives an integrand that is odd in $\kappa$. Thus, after $\kappa$ integration this term vanishes and so we have no contribution from the off-diagonal terms. So, we only have to evaluate the contribution of diagonal terms i.e. sum of Eq. \eqref{T+1_Bel1} with Eq.\eqref{T+1_Bel2} and sum of Eq. \eqref{T+2_Bel1} with Eq.\eqref{T+2_Bel2}. Belinfante total angular momentum distribution is given as:
	\begin{align}
		\langle J^z_{\text{Bel,q}} \rangle (\bsb)
		=& i \int \frac{d^2\boldsymbol{\Delta}^{\perp}}{(2\pi)^2}e^{-i\bsb \cdot \bsD}\left[ \frac{\partial \langle T^{+1}_{\text{Bel,q}}\rangle_{\text{LF}} }{\partial \Delta_{\perp}^{(2)}}-\frac{\partial \langle T^{+2}_{\text{Bel,q}}\rangle_{\text{LF}} }{\partial \Delta_{\perp}^{(1)}}\right]_{\text{DY}}.
	\end{align}
	After performing $\kappa$ integration on Eq.\eqref{T+1_Bel2} and Eq.\eqref{T+2_Bel2} and substituting them along with the single-particle contribution in the above equation, we get:
	\begin{align}
		&\nonumber \langle J^z_{\text{Bel,q}} \rangle (\bsb)\\\nonumber=& \int \frac{d^2\boldsymbol{\Delta}^{\perp}}{(2\pi)^2}e^{-i\bsb \cdot \bsD} \int dx \left\{ \left[\frac{1}{2}-\frac{g^2 C_f}{16 \pi^2} \left\{\frac{1+x^2}{1-x} \log{\left(\frac{\Lambda^2}{m^2 (1-x)^2}\right)} -\frac{2x}{1-x}\right\} \right] +\frac{g^2 C_f}{16\pi^2\left(1-x\right){\Delta}^4 \omega^3} \right. \nonumber \\ & \left. \times \bigg[\left(8m^4\left(1-2x\right)\left(1-x\left(1-x\right)\right)+6m^2\left(1-\left(2-x\right)x\left(1+2x\right)\right)\Delta^2+\left(1-\left(2-x\right)x\left(1+2x\right)\right)\Delta^4\right)\log\left(\frac{1+\omega}{-1+\omega}\right) \right.\nonumber \\& \left. -\omega \Delta^2 \left(4m^2\left(1-\left(1-x\right)x\right)+\left(1+x^2\right)\Delta^2+\left(1-\left(2-x\right)x\left(1+2x\right)\right)\left(4m^2+\Delta^2\right)\log\left(\frac{\Lambda^2}{m^2\left(1-x\right)^2}\right)\right)\bigg] \right\} .\label{JzBelq}
	\end{align}
 

\section{Spatial distribution of angular momentum components of canonical(JM) decomposition} \label{appE}
\subsection{Gluon OAM distribution}
The impact-parameter distribution of orbital angular momentum of gluon can be found from 
\begin{align}
     \langle L_{\text{can,g}}^z \rangle(\boldsymbol{b}^{\perp})
=&  i\int \frac{d^2\bsD}{(2\pi)^2}e^{-i\bsD \cdot \boldsymbol{b}^{\perp}} \left[\frac{\partial \langle T^{+1}_{\text{can,g}}\rangle_{\text{LF}} }{\partial \Delta_{\perp}^{(2)}}-\frac{\partial \langle T^{+2}_{\text{can,g}}\rangle_{\text{LF}} }{\partial \Delta_{\perp}^{(1)}}\right]_{\text{DY}},  \label{Lzcang}
\end{align}
where 
\begin{align}
    T^{+k}_{\text{can,g}}=-2\text{Tr}\left[ G^{+ \alpha } \partial^k A_\alpha \right]=G_{a}^{+\perp}~\partial^{k}A_{a}^{\perp}= \left(\partial^{+}A_{a}^{\perp}\right)\left(\partial^{k}A_{a}^{\perp}\right) 
\end{align}
The contribution to the matrix element from this term is only diagonal. The operator structure of this term is as follows 
\begin{align}
\left(\partial^{+}A_{a}^{\perp}\right)\left(\partial^{k}A_{a}^{\perp}\right) 
=& \sum_{\lambda, \lambda^{\prime}}\int \frac{dk^{+}d^2\bsk dk^{+}d^2\bskpr}{\left(16\pi^3\right)^2k^{\prime +}}k^{\prime k}\bigg[-\left(\epsilon^{\perp}_{\lambda}\cdot \epsilon^{\perp}_{\lambda^{\prime}}\right)a_{\lambda}(k)a_{\lambda^{\prime}}(k^{\prime})e^{-i\left(k+k^{\prime}\right)\cdot y} 
+ \left(\epsilon^{\perp}_{\lambda}\cdot \epsilon^{\perp*}_{\lambda^{\prime}}\right)a_{\lambda}(k)a^{\dagger}_{\lambda^{\prime}}(k^{\prime})e^{-i\left(k-k^{\prime}\right)\cdot y}  \nonumber \\ \nonumber & 
+ \left(\epsilon^{\perp*}_{\lambda}\cdot \epsilon^{\perp}_{\lambda^{\prime}}\right)a^{\dagger}_{\lambda}(k)a_{\lambda^{\prime}}(k^{\prime})e^{i\left(k-k^{\prime}\right)\cdot y} 
 -\left(\epsilon^{\perp*}_{\lambda}\cdot \epsilon^{\perp*}_{\lambda^{\prime}}\right)a^{\dagger}_{\lambda}(k)a^{\dagger}_{\lambda^{\prime}}(k^{\prime})e^{i\left(k^{\prime}+k\right)\cdot y}\bigg]
\end{align}
\begin{align}
    &\nonumber \therefore  \frac{1}{2p^{+}} \langle 2, \sigma^{\prime}|\left(\partial^{+}A_{a}^{\perp}\right)\left(\partial^{k}A_{a}^{\perp}\right) (0)|2, \sigma \rangle \\
   &= \frac{1}{2}\sum_{\lambda_{1},\lambda_{2},\lambda_{2}^{\prime}}\int dx d^2\kappa^{\perp} \phi^{\sigma^{\prime}*}_{\lambda_{1},\lambda_{2}^{\prime}}\left(x, \bs{\kappa}^{\prime\perp}\right)\bigg[\left({\kappa}^{\prime k}+x\frac{\Delta^{k}}{2}\right)\left(\bs{\epsilon}^{\perp}_{\lambda_{2}} \cdot \bs{\epsilon}^{\perp*}_{\lambda_{2}^{\prime}}\right)+\left({\kappa}^{ k}-x\frac{\Delta^{k}}{2}\right)\left(\bs{\epsilon}^{\perp*}_{\lambda_{2}^{\prime}} \cdot \bs{\epsilon}^{\perp}_{\lambda_{2}}\right)\bigg]\phi^{\sigma}_{\lambda_{1},\lambda_{2}}\left(x,\bs{\kappa}^{\perp}\right). \label{Tcan}
\end{align}
\textbf{\underline{Diagonal matrix element of $T^{\mu\nu}_{\text{can,g}}(0)$:}}\\
To evaluate the matrix element of canonical EMT, we substitute gluon LFWF expressions in Eq. \eqref{Tcan}
\begin{align}
    \nonumber &\frac{1}{2p^{+}} \langle 2, \uparrow|T^{+1}_{\text{can,g}}|2, \uparrow \rangle \\
    &= g^2 C_f\int\frac{dx d^2\boldsymbol{\kappa}^\perp}{16 \pi^3 x D_1 D_2}\left[(2 \kappa^{(1)} + (1-x) \Delta^{(1)}) (m^2 x^4 + 
   2 \kappa^{(1)} (\kappa^{(1)} +(1-x)\Delta^{(1)}) \right. \nonumber \\
   & \left.\hspace{2cm}+ x(\kappa^{(1)} + 
      i \kappa^{(2)}) (-2 + x) (\kappa^{(1)} + (1-x)\Delta^{(1)}) + (2 \kappa^{(2)} + x(-i \kappa^{(1)} + \kappa^{(2)}) (-2 + 
         x)) (\kappa^{(2)} + (1-x)\Delta^{(2)}))\right] \nonumber \\
         &= -ig^2 C_f \int \frac{dx}{16\pi^2}\, (x-2)(x-1)\Delta^{(2)}\left[-1 + \omega' \log \left(\frac{1+\omega'}{-1+\omega'} \right) - \text{log} \left(\frac{\Lambda^2}{m^2x^2} \right)\right] \label{T+k_can1} \\
         \nonumber &\frac{1}{2p^{+}} \langle 2, \uparrow|T^{+2}_{\text{can,g}}|2, \uparrow \rangle \\
    &= g^2 C_f \int\frac{dx d^2\boldsymbol{\kappa}^\perp}{16 \pi^3 x D_1 D_2}\left[(2 \kappa^{(2)} + (1-x) \Delta^{(2)}) (m^2 x^4 + 
   2 \kappa^{(1)} (\kappa^{(1)} +(1-x)\Delta^{(1)}) \right. \nonumber \\
   & \left.\hspace{2cm}+ x(\kappa^{(1)} + 
      i \kappa^{(2)}) (-2 + x) (\kappa^{(1)} + (1-x)\Delta^{(1)}) + (2 \kappa^{(2)} + x(-i \kappa^{(1)} + \kappa^{(2)}) (-2 + 
         x)) (\kappa^{(2)} + (1-x)\Delta^{(2)}))\right]\nonumber \\
         &= ig^2 C_f \int \frac{dx}{16\pi^2}\, (x-2)(x-1)\Delta^{(1)}\left[-1 + \omega' \text{log} \left(\frac{1+\omega'}{-1+\omega'} \right) - \text{log} \left(\frac{\Lambda^2}{m^2x^2} \right)\right] \label{T+k_can2}
\end{align} 
Substituting these components in Eq. \eqref{Lzcang}, we get
\begin{align}
    &\langle L_{\text{can,g}}^z \rangle(\boldsymbol{b}^{\perp}) \nonumber \\
    &= g^2 C_f \int \frac{d^2\boldsymbol{\Delta}^{\perp}}{(2\pi)^2}e^{-i\bsb \cdot \bsD} \int \frac{dx}{8 \pi^2} \frac{x-2}{(x-1){\Delta}^2 \omega'} \left[ (2 m^2 x^2 + (x-1)^2 \Delta^2) \text{log} \left( \frac{1 + \omega'}{-1 + \omega'} \right) - (x-1)^2 \Delta^2 \omega' \text{log} \left( \frac{\Lambda^2}{m^2 x^2} \right)
 \right]
\end{align}

\subsection{Gluon spin distribution}
The impact-parameter distribution of spin angular momentum of gluon can be found from
\begin{align}
     \langle S_{\text{can,g}}^z \rangle(\boldsymbol{b}^{\perp})
= \left. \frac{1}{2}\epsilon^{3jk}\int \frac{d^2\bsD}{(2\pi)^2}e^{-i\bsD \cdot \boldsymbol{b}^{\perp}} \langle S^{+jk}_{\text{can,g}}\rangle \right \vert_{\text{DY}}, \label{Sz_dist}
\end{align}
where, 
\begin{align}
    S^{\mu\nu\rho}_{\text{can,g}}=- 2\text{Tr}\left[ G^{\mu [ \nu} A^{\rho]} \right]
\end{align}
The operator structure corresponding to the gluon spin in the canonical decomposition can be expressed as 
\begin{align}
   \nonumber S^{+jk}_{\text{can,g}}=&- 2\text{Tr}\left[ G^{+ [ j} A^{k]} \right]=-G^{+j}_{a}A^{k}_{a}+G^{+k}_{a}A^{j}_{a}=-\left(\partial^{+}A_{a}^{j}\right)A_{a}^{k}+\left(\partial^{+}A_{a}^{k}\right)A_{a}^{j}\\ \nonumber =&i\sum_{\lambda,\lambda^{\prime}}\int \frac{dk^{+}d^2\bs{k}^{\perp} dk^{\prime +}d^2\bs{k}^{\perp}}{\left(16\pi^3\right)^{2}k^{\prime +}}\bigg[\left(\epsilon^{j}_{\lambda}\epsilon^{k}_{\lambda^{\prime}}-\epsilon^{k}_{\lambda}\epsilon^{j}_{\lambda^{\prime}}\right)a_{\lambda}(k)a_{\lambda^{\prime}}(k^{\prime})e^{-i\left(k+k^{\prime}\right)\cdot y}+\left(\epsilon^{j}_{\lambda}\epsilon^{k*}_{\lambda^{\prime}}-\epsilon^{k}_{\lambda}\epsilon^{j*}_{\lambda^{\prime}}\right)a_{\lambda}(k)a^{\dagger}_{\lambda^{\prime}}(k^{\prime})e^{-i\left(k-k^{\prime}\right)\cdot y}\\ & -\left(\epsilon^{j*}_{\lambda}\epsilon^{k}_{\lambda^{\prime}}-\epsilon^{k*}_{\lambda}\epsilon^{j}_{\lambda^{\prime}}\right)a^{\dagger}_{\lambda}(k)a_{\lambda^{\prime}}(k^{\prime})e^{i\left(k-k^{\prime}\right)\cdot y}-\left(\epsilon^{j*}_{\lambda}\epsilon^{k*}_{\lambda^{\prime}}-\epsilon^{k*}_{\lambda}\epsilon^{j*}_{\lambda^{\prime}}\right)a^{\dagger}_{\lambda}(k)a^{\dagger}_{\lambda^{\prime}}(k^{\prime})e^{i\left(k+k^{\prime}\right)\cdot y}\bigg]
\end{align}
The contribution to the matrix element from this operator is also purely diagonal 
\begin{align}
\nonumber    &\frac{1}{2p^{+}}\langle 2, \sigma^{\prime}|-\left(\partial^{+}A_{a}^{j}\right)A_{a}^{k}+\left(\partial^{+}A_{a}^{k}\right)A_{a}^{j}|2,\sigma \rangle \\ =& \frac{i}{2}\sum_{\lambda_{1},\lambda_{2},\lambda_{2}^{\prime}}\int dx d^2\bska \phi^{\sigma^{\prime}*}_{\lambda_{1},\lambda_{2}^{\prime}}\left(x,\bskapr\right)\left[\left(\epsilon^{j}_{\lambda_{2}}\epsilon^{k*}_{\lambda_{2}^{\prime}}-\epsilon^{k}_{\lambda_{2}}\epsilon^{j*}_{\lambda_{2}^{\prime}}\right)-\left(\epsilon^{j*}_{\lambda_{2}^{\prime}}\epsilon^{k}_{\lambda_{2}}-\epsilon^{k*}_{\lambda_{2}^{\prime}}\epsilon^{j}_{\lambda_{2}}\right)\right]\phi^{\sigma}_{\lambda_{1},\lambda_{2}}\left(x,\bska\right)
\end{align}
Substituting this matrix element in the Eq. \eqref{Sz_dist}, we get 
\begin{align}
  \langle S_{\text{can,g}}^z &\rangle(\boldsymbol{b}^{\perp})
= g^2 C_f \int \frac{d^2\boldsymbol{\Delta}^{\perp}}{(2\pi)^2}e^{-i\bsb \cdot \bsD} \int \frac{dx}{{8 \pi^2 \Delta^4}} \times \nonumber  \\ &\left[ \frac{{((x-2 ) (x-1)^2 \Delta^4 + 2 m^2 x^2 ((3x-4) \Delta^2 - (x-2) \Delta^2)) \log\left(\frac{{1 + \omega'}}{{-1 + \omega'}}\right)}}{{(x-1)^2 \omega'}} - (x-2) \Delta^4 \log\left(\frac{{\Lambda^2}}{{m^2 x^2}}\right) \right]
\end{align}

\subsection{Spatial distribution of the superpotential for gluon}
Using the expression given in Eq. \eqref{Mzgeneral} with $\kappa=1$ for gluon superpotential and the form of canonical spin angular momentum operator given in the previous subsection, we get
\begin{align}
   & \langle M^{z}_{g}\rangle (\bsb) = \epsilon^{3jk} \int \frac{d^2 \bsD}{(2\pi)^{2}}~e^{-i\bsb \cdot \bsD}  ~\Delta^{l}\frac{\partial }{\partial \Delta^{j}} \langle G^{+l}A^{k}  \rangle \nonumber \\  =&  \int \frac{d^2 \bsD}{(2\pi)^{2}}~e^{-i\bsb \cdot \bsD}  ~ \left[\Delta^{(1)}\left(\frac{\partial }{\partial \Delta^{(1)}} \langle G^{+1}A^{(2)}\rangle - \frac{\partial }{\partial \Delta^{(2)}} \langle G^{+1}A^{(1)}\rangle \right) + \Delta^{(2)}\left(\frac{\partial }{\partial \Delta^{(1)}} \langle G^{+2}A^{(2)}\rangle - \frac{\partial }{\partial \Delta^{(2)}} \langle G^{+2}A^{(1)}\rangle \right) \right] \nonumber 
   \\ =&  \int \frac{d^2 \bsD}{(2\pi)^{2}}~e^{-i\bsb \cdot \bsD} ~ \left[\Delta^{(1)}\left(\frac{\partial }{\partial \Delta^{(1)}} \langle G^{+1}A^{(2)}\rangle \right) - \Delta^{(2)}\left(  \frac{\partial }{\partial \Delta^{(2)}} \langle G^{+2}A^{(1)}\rangle \right) \right] \label{Mzgluon}
\end{align}
where, $\langle G^{+l}A^{k}  \rangle = \frac{\langle p^{\prime}, \bs{s} |G^{+l}A^{k} |p, \bs{s} \rangle}{2p^{+}}$.
The structure of the corresponding operator 
\begin{align}
\nonumber    G^{+l}A^{k} =  \left(\partial^{+}A^{l}\right)A^{k} \nonumber =& i \sum_{\lambda, \lambda^{\prime}}\int \frac{dk^{\prime +}d^{2}\bskpr dk^{+} d^{2}\bsk}{\left(16 \pi^{3}\right)^{2} k^{\prime +}} \bigg[- \epsilon_{\lambda}^{l} \epsilon_{\lambda^{\prime}}^{k} a_{\lambda}(k) a_{\lambda^{\prime}}(k) e^{-i \left(k+k^{\prime}\right) \cdot y } - \epsilon_{\lambda}^{l} \epsilon_{\lambda^{\prime}}^{*k} a_{\lambda}(k) a^{\dagger}_{\lambda^{\prime}}(k) e^{-i \left(k-k^{\prime}\right) \cdot y } \\& + \epsilon_{\lambda}^{*l} \epsilon_{\lambda^{\prime}}^{k} a^{\dagger}_{\lambda}(k) a_{\lambda^{\prime}}(k) e^{-i \left(k^{\prime}-k\right) \cdot y } + \epsilon_{\lambda}^{*l} \epsilon_{\lambda^{\prime}}^{*k} a^{\dagger}_{\lambda}(k) a^{\dagger}_{\lambda^{\prime}}(k) e^{i \left(k+k^{\prime}\right) \cdot y } \bigg]
\end{align}
and the matrix element of this operator is given by 
\begin{align}
   & \frac{\langle p^{\prime}, \bs{\sigma} |G^{+l}A^{k}(0)| p, \bs{\sigma} \rangle}{2p^{+}} = \frac{i}{2} \sum_{\lambda_{1}, \lambda_{2}, \lambda_{2}^{\prime}} \int dx d^2 \bska \phi^{\bs{s}*}_{\lambda_{1}, \lambda_{2}^{\prime}}\left(x, \bskapr \right)\left[- \epsilon^{l}_{\lambda_{2}} \epsilon^{*k}_{\lambda_{2}^{\prime}} + \epsilon^{*l}_{\lambda_{2}^{\prime}} \epsilon^{k}_{\lambda_{2}} \right] \phi^{\bs{s}}_{\lambda_1, \lambda_{2}}\left(x, \bska\right) 
\end{align}
We now evaluate the individual components by substituting the LFWF and then doing the $\kappa$ integrations 
\begin{align}
    &\frac{\langle p^{\prime}, \bs{\sigma} |G^{+1}A^{2}| p, \bs{\sigma} \rangle}{2p^{+}} \nonumber \\
    &= g^2 C_f \int \frac{dx}{16\pi^2 \Delta^4} \left[\frac{{(-(x - 2) (x - 1)^2 \Delta^4 + 2 m^2 x^2 ((4 - 3 x) \Delta^2 + (x - 2) \Delta^2)) \log\left(\frac{{1 + \omega'}}{{-1 + \omega'}}\right)}}{{(x - 1)^2 \omega'}} + (x - 2) \Delta^4 \log\left(\frac{{\Lambda^2}}{{m^2 x^2}}\right)
\right]  \label{S1+2}\\
&\frac{\langle p^{\prime}, \bs{\sigma} |G^{+2}A^{1}| p, \bs{\sigma} \rangle}{2p^{+}} \nonumber \\
&= g^2 C_f \int \frac{dx}{16\pi^2 \Delta^4} \left[\frac{{((x - 2) (x - 1)^2 \Delta^4 + 2 m^2 x^2 (( 3 x-4) \Delta^2 -(x - 2) \Delta^2)) \log\left(\frac{{1 + \omega'}}{{-1 + \omega'}}\right)}}{{(x - 1)^2 \omega'}} - (x - 2) \Delta^4 \log\left(\frac{{\Lambda^2}}{{m^2 x^2}}\right)\right] \label{S2+1}
\end{align}
Substituting these in Eq. \eqref{Mzgluon}, we get the impact-parameter distribution of the correction term 
\begin{align}
    \langle M^{z}_{g}\rangle (\bsb) =& g^2 C_f \int \frac{d^2\boldsymbol{\Delta}^{\perp}}{(2\pi)^2}e^{-i\bsb \cdot \bsD} \int \frac{dx}{8\pi^2 (1-x)^3\Delta^4 \omega'^3} \times \nonumber \\
    &\left[2 m^2 x^3 (4 m^2 x + (x - 1) \Delta^2) \log\left(\frac{{1 + \omega'}}{{-1 + \omega'}}\right) -  \Delta^2 \omega'(x - 1)^2 (4 m^2 x^2 + (x - 2) (x - 1) \Delta^2)
\right] \hspace{1 cm}
\end{align}

	\bibliographystyle{elsarticle-num}
	\bibliography{references}	
\end{document}